\documentclass[api,letterpaper,onecolumn,groupedaddress,footinbib,amssymb,longbibliography]{revtex4}
\usepackage[utf8x]{inputenc}
\usepackage[xdvi]{graphicx}
\usepackage[english]{babel}
\usepackage{amsmath}
\usepackage{amssymb}
\usepackage{graphicx}
\usepackage{xcolor}
\usepackage{enumerate}
\usepackage[colorlinks=true, linkcolor=black, filecolor=blue, urlcolor=blue, citecolor=blue]{hyperref}

\begin{document}

\title{From Molecular Quantum Electrodynamics at Finite Temperatures\\ to Nuclear Magnetic Resonance.}

\author{Kolja Them$^{1,2,3}$  \\ \medskip 
$^1$\footnotesize Section Biomedical Imaging, Molecular Imaging North Competence Center, Kiel University, Am Botanischen Garten 14, 24118 Kiel, Germany \\ \smallskip
$^2$ Department of Radiology and Neuroradiology, University Medical Center Schleswig-Holstein, Arnold-Heller-Stra{\ss}e 3, 24105 Kiel, Germany \\ \smallskip
$^3$ kolja.them@rad.uni-kiel.de
}

\begin{abstract} 
\textbf{Abstract.}
The algebraic reformulation of molecular Quantum Electrodynamics (mQED) at finite temperatures is applied to Nuclear Magnetic Resonance (NMR) in order to provide a foundation for the reconstruction of much more detailed molecular structures, than possible with current methods. Conventional NMR theories are based on the effective spin model which idealizes nuclei as fixed point particles in a lattice $L$, while molecular vibrations, bond rotations and proton exchange cause a delocalization of nuclei. Hence, a lot information on molecular structures remain hidden in experimental NMR data, if the effective spin model is used for the investigation.
\medskip \\
In this document it is shown how the quantum mechanical probability density $\mid\Psi^\beta(X)\mid^2$ on $\mathbb{R}^{3n}$ for the continuous, spatial distribution of $n$ nuclei can be reconstructed from NMR data. To this end, it is shown how NMR spectra can be calculated directly from mQED at finite temperatures without involving the effective description. The fundamental problem of performing numerical calculations with the infinite-dimensional radiation field is solved by using a purified representation of a KMS state on a $W^*$-dynamical system. Furthermore, it is shown that the presented method corrects wrong predictions of the effective spin model. It is outlined that the presented method can be applied to any molecular system whose electronic ground state can be calculated using a common quantum chemical method. Therefore, the presented method may replace the effective spin model which forms the basis for NMR theory since 1950.  
\end{abstract}

\maketitle

\tableofcontents   

\section{Introduction} \label{Sec:Introduction}
Advances in chemistry, pharmacy, structure-based drug design and nanoscience often depend on the detailed knowledge of a molecular structure, which is determined by the spatial distribution of the nuclei \cite{StructureI}. In particular, the pharmacological properties of drugs depend heavily on small details of the charge distribution in the molecular structure \cite{StructureIII}. In 1946, the experimental technique of Nuclear Magnetic Resonance (NMR) spectroscopy was developed, which is nowadays one of the most used and most advanced methods for molecular structure determination  \cite{NMR, NMR2}. NMR data contain highly detailed information on the spatial distribution of the nuclei including binding lengths, binding angles, bond rotations, molecular vibrations, proton exchange and the electronic influence of neighboring molecules \cite{Rotation}. From 1950 to 1953 Norman Ramsey calculated the chemical shift observed in NMR from the energy of the ground state, eq. (\ref{eq:Ramsey}), and thus laid a foundation for today's NMR theory for structural analysis \cite{holzschuh2016quantenmechanische, NMRTheory, NMRTheory2, NMRTheory3}. However, even 74 years after the invention of NMR, much information about molecular structures cannot be decoded and remains hidden in experimental NMR data. This concerns especially molecular structures where the positions of the nuclei cannot adequately be described as fixed points in space. 
Due to the finite temperature the nuclei of a molecule are generally distributed in space to all positions which are accessible through thermal energy. Important examples for such situations are bond rotations and vibrations which can lead to interconversions and superpositions of different conformations of a molecule. If Classical Physics is used to describe bond rotations, the nuclei rotate with a certain frequency and hence have time-dependent positions. In conventional NMR theory, this concept is phenomenologically described in the form of rate constants describing the rotation frequency. 
However, in the more realistic theory of Quantum Statistical Mechanics bond rotations are included in wave functions $\Psi^\beta$ whose amplitude square $\mid\Psi^\beta(X)\mid^2$ provides the \textit{continuous probability distribution} to find the nuclei with conformation $X$. Thus, the temperature has an important impact on the molecular structure and causes a delocalization of certain nuclei. Such effects can often be observed in NMR spectra \cite{cordier2002temperature}. The description of such delocalized nuclei using a spatial probability distribution $\mid\Psi^\beta(X) \mid^2$ is obviously more realistic and more detailed compared to an idealization as fixed point particles in combination with phenomenological rate constants.

\begin{figure}[h]     
\centering
\includegraphics*[scale=0.9]{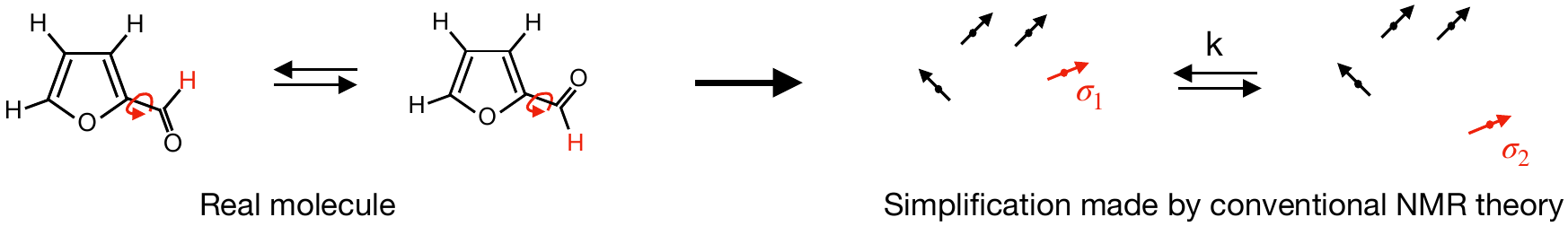}
\caption{\label{fig:1} Conventional NMR theory makes too much simplifications such that occupation probabilities for bond angles remain hidden in experimental NMR data. Instead, the molecule is simplified with two different structures with fixed points for the positions of the nuclei and a rate constant $k$ describing the time required for interconversion of these two structures. The idea, that the bond (marked by the curved red arrow) is rotating with a certain frequency, is based on Classical Physics. However, the more realistic situation in thermal equilibrium is that each bond angle $\theta$ is occupied with a certain probability $\mid \Psi^\beta(\theta)\mid^2$.}
\end{figure}

However, conventional NMR theory is based on the effective spin model, eq. (\ref{eq:Eff}), which idealizes nuclei as point particles at fixed positions $x_i$ and whose thermal states are almost independent from the temperature \cite{holzschuh2016quantenmechanische, NMRTheory, NMRTheory2, NMRTheory3}. The effective spin model had certainly great success over the last decades \cite{pravdivtsev2019simulating}, but it also suffers from the fact that delocalization of nuclei due to bond rotations, vibrations and proton exchange can only be included phenomenologically \cite{PhenoSpin, PhenoSpinII}. 
The phenomenological description in the form of rate constants gives a rough insight into these effects \cite{PhenoSpinIII}, but it also prevents a desirable analysis of the more realistic, continuous probability density $\mid\Psi^\beta(X)\mid^2$ for the spatial distribution of the nuclei. The effective model \cite{edwards2014quantum, bak2011simpson} 

\begin{equation}\label{eq:Eff}
  H_{\mathrm{eff}}=-\sum_i\gamma_i\vec{I}_i(\hat{1}-\sigma_i)\vec{B}_\mathrm{ext}+2\pi\sum_{i< j}J_{ij}\vec{I}_i\cdot\vec{I}_j+\sum_{i<j}\vec{I}_iD_{ij}\vec{I}_j,
\end{equation} 
contains the magnetic shielding $\sigma_i$, which is caused by surrounding electrons. The magnetic moments of these electrons show into the opposite direction of the external magnetic field and hence weakens the external field at the position of a nucleus. The indirect spin-spin couplings $J_{ij}$ are also caused by electrons and enable energy exchange between nuclei at $i$ and $j$. The tensor $D_{ij}$ describes the magnetic dipole-dipole interactions between the nuclear spins $\vec{I}_i$ and $\vec{I}_j$, $\vec{B}_\mathrm{ext}$ is a classical, external magnetic field and $\gamma$ is the gyromagnetic ratio  \cite{holzschuh2016quantenmechanische, helgaker2000analytical}. 
 In the most widely used approach the effective parameters are calculated according to second order derivatives of the ground state energy (Taylor-expansion) \cite{holzschuh2016quantenmechanische, NMRTheory, NMRTheory2, NMRTheory3, NMRTheory4}:
\begin{equation}\label{eq:Ramsey}
(\underline{\sigma}_i)_{\alpha\beta}=\frac{\partial^2 E_0}{\partial \mu_{i}^\alpha\partial B^\beta} \quad\quad \mathrm{and} \quad\quad (\underline{J}_{ij})_{\alpha\beta}=h\gamma_i\gamma_j \frac{d^2 E_0}{d\mu_{i}^\alpha d\mu_{j}^\beta}.
\end{equation}
 During the last decades there were done many works on the optimization of eq. (\ref{eq:Ramsey}) by including relativistic \cite{autschbach2009relativistic, demissie2017relativistic, cheng2009four} and QED effects \cite{romero2002qed, romero2002self, yerokhin2011qed, gimenez2016quantum} to the effective NMR parameters. Also numerically more efficient alternatives were introduced \cite{aucar2010polarization, aucar2008understanding}.  However, all these methods provide parameters for the effective model which restricts the positions of the nuclei to fixed points in a lattice $L$ and which possesses a \textit{discrete energy spectrum}. 
 In contrast, signals in NMR spectra are continuous and have an important width and shape, which is due to the process of \textit{return to equilibrium} (relaxation or thermalization) and also due to delocalization of nuclei. Thus, a \textit{continuous} spectrum is observed in NMR and a discrete spectrum can only be an approximation for very narrow shaped signals. Furthermore, the unitary dynamics generated by (\ref{eq:Eff}) has bad thermalization properties. Small systems consisting of a few spins does not thermalize at all and larger systems thermalize only approximately in very specific cases \cite{Them}. In order to include return to equilibrium anyway \cite{farraher2006differentiation} the von Neumann equation was modified phenomenologically by introducing relaxation superoperators $\Gamma$ \cite{kuprov2011diagonalization, kleier1970general}: 
\begin{equation}\label{eq:effRelax}
\frac{d \rho(t)}{dt}=-\frac{i}{\hbar}[H_\mathrm{eff},\rho(t)]-\Gamma(\rho(t)-\rho_0).
\end{equation} The final state $\rho_0$ to which the system shall evolve must be chosen "by hand". Certainly, it is preferable when the correct final state is an outcome and not an input of a theory. Conventional NMR methods to study certain line shapes caused by effects such as molecular rotations or proton exchange uses even more phenomenology. These methods are frequently used in dynamic NMR and the line shape analysis is often based on the Bloch-McConnell equations or related methods \cite{lounila1984effects, LineExchange, LineExchangeII, LineExchangeIII}. In all these methods, the microscopic origin of NMR line shapes is completely neglected and replaced by phenomenological parameters like rate constants $k$ or relaxation parameters. This is due to the inability of the effective model to use a continuous, spatial distribution to describe delocalized nuclei and because of the discrete energy spectrum of eq. (\ref{eq:Eff}). An illustrative example where conventional NMR methods fail is the investigation of the probability distribution for the occupation of certain bond angles (fig. \ref{fig:1}). In the real molecule (which is furfural in this example) all bond angles can be occupied via bond rotations. We know from quantum mechanics that every bond angle has a specific energy and the laws of thermodynamics provide the information about which of these bond angles are preferably occupied. Conventional NMR theory like the Bloch-McConnell equations, however, simplifies the molecule usually by using two different structures with fixed point positions for the nuclei and a rate constant $k$ (right side in fig. \ref{fig:1}). This rate constant describes the time required for the mutual conversion of these structures. Hence, the important information which bond angles are more and which are less preferred at a given temperature cannot be decoded from experimental NMR data by using conventional NMR theory.  

In this document it is shown how the probability distribution $\mid \Psi^\beta(X)\mid^2$ for the positions $X$ of the nuclei can be analyzed and reconstructed from NMR spectra. 
To this end, it is shown how the NMR signal can be calculated \textit{directly} from molecular Quantum Electrodynamics (mQED) at finite temperatures without using the effective spin model or effective NMR parameters. Hence, the presented method can be used to obtain a more detailed molecular structure from NMR data than currently possible with conventional NMR theory. Mathematically, this means that the lattice $L$, which serves for the restricted positions of the nuclei in the effective description, is replaced by the continuous space $\mathbb{R}^3$ in which the nuclei can be distributed continuously. A reconstruction of $\mid \Psi^\beta(X)\mid^2$ from NMR data is of special interest, because in most cases it is not possible to solve the nuclear Schr\"{o}dinger equation. An outlook how a more detailed structure determination may look like is presented in section \ref{sec:Outlook}.

However, up to now it was not known how the spin dynamics can be calculated numerically when the spins interact with the infinitely dimensional, quantized electromagnetic field (EMF) with a \textit{continuous} spectrum at finite temperatures. Two basic reasons for that are the occurrence of divergences in perturbation series and the infinite number of field quanta involved in finite temperature QED processes. In certain cases one may avoid the numerical and mathematical problems related to quantized fields at finite temperatures by using the ground state instead \cite{ruggenthaler2018quantum, schafer2018ab}. However, in NMR at room temperature the nuclear spins are far away from their ground state and the temperature of the quantized electromagnetic field determines the temperature of the final state of the nuclear spins after equilibration \cite{muck2005thermal}. Hence, the approach of using a ground state for the quantized EMF is obviously unsuitable for NMR at room temperature. There are several works on a method called Thermo Field Dynamics (TFD) \cite{santana2000symmetry, ojima1981gauge, santana1999w} which is about quantized fields at finite temperatures. While this approach is widely used it also involves a large number of field quanta in the construction of the thermal state $\mid O(\beta)\rangle$:
\begin{equation}\label{eq:TFD}
\mid O(\beta)\rangle=Z(\beta)^{-1/2}\sum_n e^{-\beta E_n/2}\mid n ,\tilde{n}\rangle \quad \mathrm{or} \quad
\mid O(\beta)\rangle=  Z(\beta)^{-1/2}\sum_n  \frac{e^{-\beta E_n/2}}{n!}(a^\dagger)^n(\tilde{a}^\dagger)^n\mid 0,\tilde{0}\rangle.
\end{equation}
$\mid n ,\tilde{n}\rangle$ is an Eigenstate of the Hamiltonian, $a^\dagger$ and $\tilde{a}^\dagger$ are creation operators and $\mid 0,\tilde{0}\rangle$ is a thermal vacuum state. Hence, the calculation of an expectation value in this state involves a relatively large number of photons in the numerical calculations. Furthermore, the state $\mid O(\beta)\rangle$ in eq. (\ref{eq:TFD}) is constructed using a \textit{discrete} set of energy values $E_n$. An extension of eq. (\ref{eq:TFD}) to the continuous case is not possible. 
The use of a discrete energy spectrum for the quantized EMF and the limitation to a few (usually 1 - 100) frequencies is the common approximation made in current numerical methods. The discrete spectrum drastically simplifies the mathematical structure. No numerical method was found in the literature that uses a continuous energy spectrum for the quantized EMF to calculate the spin dynamics. However, as it turns out later in this document a discrete spectrum does not lead to satisfying results in the calculation of NMR spectra. Indeed, the incorporation of a continuous spectrum for the quantized EMF is of paramount importance for the NMR line width which is directly related with return to equilibrium properties and determines the life time of excited spins. Hence, TFD is unsuitable for the calculation of NMR spectra. 

In the present document the following problems for numerical methods are solved by using the mathematical structure from algebraic Quantum Field Theory \cite{BratteliRobinsonI, BratteliRobinsonII, aQFTI, muck2005thermal, fredenhagen2014construction, aQFTII, aQFTIII}:
\begin{enumerate}[(I)]
\item Numerical calculations with the infinite-dimensional, quantized EMF at finite temperatures.
\item Numerical calculations with a continuous energy spectrum for the quantized EMF.
\item Convergence of the QED perturbation series.
\end{enumerate}
Recent works investigated and avoided the occurrence of divergences by using appropriate smearing functions \cite{amour2015hamiltonian, amour2017quantum, muck2005thermal}. It remained to show which effect these restrictions have on expectation values, which will be done in this work. Recently, a  perturbation series for interacting, massive quantum fields was constructed by Fredenhagen and Lindner  \cite{fredenhagen2014construction}. This approach solved a long-standing problem and its extension to the Dirac field is of interest for relativistic effects from heavy nuclei in NMR. Further important structural developments were achieved in \cite{drago2017generalised}. 
In this document it is shown that a purified form of the \textit{Araki-Woods representation} \cite{araki1963representations}, denoted by $(\mathfrak{H}_\mathrm{AW}, \pi_\mathrm{AW}^\beta)$, enables the numerical calculations involving bosonic fields at finite temperatures with striking advantages: In each order of the perturbation series at most one "Araki-Woods boson" is produced while small coupling constants, connecting spins and the quantized electromagnetic field, reduce higher order contributions. The representation $(\mathfrak{H}_\mathrm{AW}, \pi_\mathrm{AW}^{\beta})$ rigorously respects the continuous energy spectrum of the quantized electromagnetic field at finite temperatures and reduces the required computational resources for numerical calculations strongly. Furthermore, this document shows that the application of mQED to NMR in the algebraic reformulation of Quantum Field Theory offers the following advantages over conventional NMR theory: 
\begin{enumerate}[(I)]
\item The drawback of a nearly temperature-independent initial state from which conventional NMR theory suffers (effective spin model) can be repealed. Instead, a suitable, temperature-dependent probability density for the spatial distribution of the nuclei can be used. This basically enables a much more detailed reconstruction of the molecular structures contained in NMR data.      
\item There is a direct and causal connection between NMR line shapes and molecular structures. Hence, no phenomenological parameters prevent the reconstruction of the spatial distribution of delocalized nuclei.
\item Molecular rotations, vibrations and proton exchange can be included in the probability density for the spatial distribution of the nuclei. Hence, the simplification that the positions of nuclei are restricted to fixed points can be repealed. 
\item Thermal equilibration is naturally contained in the unitary QED spin-dynamics \cite{derezinski2003return, muck2005thermal}. 
\end{enumerate}

\section{Molecular Quantum Electrodynamics}
In order to use the perturbation theory developed by Araki, Bratelli, Robinson and Kishimoto the Hamiltonian will be separated into $H=H_0+H_{\mathrm{Int}}$. The physical system will be described by a combination of a Pauli-Fierz and a generalized Spin-Bose model in Coulomb gauge \cite{derezinski2003return, amour2015hamiltonian}. The resulting molecular QED Hamiltonian is given by
\begin{equation}\label{eq:FreeQED}
H_0=-\big(\sum_{j=1}^K \gamma_j\vec{I}_j +\sum_{i=1}^E \vec{\mu_{i\mathrm{J}}}\big)\cdot\vec{B}_\mathrm{ext} 
+\hbar\sum_{\lambda=1,2}\int_{\mathbb{R}^3}
d^3k\omega(\vec{k})a^*(\vec{k},\lambda)a(\vec{k},\lambda)+\sum_{i=1}^E\frac{\vec{p_i}^2}{2m_\mathrm{e}}
+\sum_{j=1}^K\frac{\vec{P_j}^2}{2M_j}  +V(X^\mathrm{e},X)
\end{equation}
and
\begin{equation}\label{eq:IntQED}
H_{\mathrm{Int}}=-\sum_{j=1}^K \gamma_j\vec{I}_j\cdot\vec{B}_\varphi(\vec{x_j})+ \sum_{i=1}^E\bigg(i\frac{e\hbar}{m_\mathrm{e}}\vec{A}_\varphi(\vec{x_i^\mathrm{e}})\cdot\vec{\nabla}^\mathrm{e}_i+\frac{e^2}{2m_\mathrm{e}}\big(\vec{A}_\varphi(\vec{x_i^\mathrm{e}})\big)^2-\vec{\mu_{i\mathrm{J}}}\cdot\vec{B}_\varphi(\vec{x^\mathrm{e}_i})\bigg).
\end{equation} The first term in $H_0$ couples the $K$ nuclear spins $\vec{I}_j$ and the $E$ total magnetic moments $\vec{\mu_{i\mathrm{J}}}=-\mu_\mathrm{B}/\hbar(g_\mathrm{e}\vec{s}_i+\vec{l}_i)$ of the electrons to the classical, external magnetic field $\vec{B}_\mathrm{ext}$. For high field strengths of the external magnetic field, i.e., $B^\mathrm{z}_{\mathrm{ext}}>3\mathrm{T}$, spin-orbit couplings can be neglected due to the Paschen-Back effect. The second term describes the energy of the quantized, electromagnetic field. $a^*(\vec{k},\lambda)$  and $a(\vec{k},\lambda) $  are the common creation and annihilation operators with commutation relations
\begin{equation}\label{eq:commutation}
[a(\vec{k},\lambda),a^*(\vec{k'},\lambda')]=\delta(\vec{k}-\vec{k}')\delta_{\lambda,\lambda'}, \quad [a(\vec{k},\lambda),a(\vec{k'},\lambda')]=0, \quad \mathrm{and} \quad[a^*(\vec{k},\lambda),a^*(\vec{k'},\lambda')]=0
\end{equation} and with momentum $\vec{k}\in\mathbb{R}^3$ and polarization $\lambda=1,2$. The last three terms provide the non-relativistic Schr\"{o}dinger-Operator. Thus, $\vec{p}_i$ is the momentum operator of electron \textit{i}, $\vec{P}_j$  is the momentum operator of nucleus \textit{j} and the potential $V(X^\mathrm{e},X)$ depending on coordinates $X^\mathrm{e}$ of \textit{E} electrons and $X$ of \textit{K} nuclei is given by
\begin{equation}
V(X^\mathrm{e},X)=\sum_{i<j}^E\frac{e^2}{4\pi\epsilon_0 x^\mathrm{e}_{ij}}-\sum_{i=1}^E\sum_{j=1}^K\frac{Z_j e^2}{4\pi\epsilon_0\parallel \vec{x^\mathrm{e}_i}-\vec{x_j}\parallel}+\sum_{i<j}^K\frac{Z_i Z_j e^2}{4\pi\epsilon_0 x_{ij}}.
\end{equation} $Z_j$ is the number of protons in nucleus $j$, $\vec{x^\mathrm{e}_i}$ is the coordinate of electron $i$ and $\vec{x}_j$ of nucleus $j$. $x_{ij}^\mathrm{e}$ and $x_{ij}$ are the distances between electrons or nuclei and the other constants can be found in the literature \cite{lieb2005bound}. $H_{\mathrm{Int}}$ couples the independent terms and enables energy exchange between the nuclear spins and the rest of the system. We use the definition $\vec{A}_{\varphi}(\vec{x_j})\dot{=}\vec{A}_{0\varphi}(\vec{x_j},0)$ and for the quantized vector potential the free time evolution provides
\begin{equation}
\vec{A}_{0\varphi}(\vec{x},t)=\sqrt{\frac{\hbar}{2\varepsilon_0 (2\pi)^3}}\sum_{\lambda=1,2}\int_{\mathbb{R}^3}d^3k\vec{\epsilon}_\lambda(\vec{k})\frac{\varphi(\vec{k})}{\sqrt{\omega(\vec{k})}}\bigg(e^{-i(\vec{k}\vec{x}-\omega(\vec{k})t))}a^*_\lambda(\vec{k}) +e^{i(\vec{k}\vec{x}-\omega(\vec{k})t)}a_\lambda(\vec{k})\bigg).
\end{equation} The quantized magnetic field is given by $\vec{B}_\varphi=\vec{\nabla}\times\vec{A}_\varphi$ and we have
\begin{equation}\label{eq:MagneticField}
\vec{B}_{0\varphi}(\vec{x},t)=i\sqrt{\frac{\hbar}{2\epsilon_0(2\pi)^3}}\sum_{\lambda=1,2}\int_{\mathbb{R}^3}d^3k\big(\vec{k}\times\vec{\epsilon}_\lambda(\vec{k})\big)\frac{\varphi(\vec{k})}{\sqrt{\omega(\vec{k})}}\bigg(e^{-i(\vec{k}\vec{x}-\omega(\vec{k})t)}a^*_\lambda(\vec{k}) -e^{i(\vec{k}\vec{x}-\omega(\vec{k})t)}a_\lambda(\vec{k})\bigg),
\end{equation} where $\varphi\in L^2(\mathbb{R}^3)$ is the coupling function with suitable IR and UV behavior \cite{derezinski2003return, lieb2005bound, lieb2004note} to prevent divergences in the individual terms of the perturbation series. The presented model is independent of a specific choice of the polarization vectors. Using the notation $x=(\vec{x},t), \, k=(\vec{k},\omega(\vec{k}))$ and Einstein's sum convention $k^\mu x_\mu=\vec{k}\cdot\vec{x}-\omega(\vec{k})t$ we have 
\begin{equation}
[B^\alpha_{0\varphi}(x),B^\gamma_{0\varphi}(y)]=-\sum_{\lambda=1,2}\,\int d^3k\varphi^\alpha_\lambda(\vec{k})\varphi^\gamma_\lambda(\vec{k})\,i\Delta_{\vec{k}}(x-y) 
\end{equation}
with $\varphi^\alpha_\lambda(\vec{k})\dot{=}(\sqrt{\hbar/\epsilon_0})\varphi(\vec{k})(\vec{k}\times\vec{\epsilon_\lambda}(\vec{k}))^\alpha$, $\alpha,\gamma=\mathrm{x},\mathrm{y},\mathrm{z}$ and
\begin{equation}
i\Delta_{\vec{k}}(x-y)=\frac{e^{-ik^\mu(x_\mu-y_\mu)}-e^{ik^\mu(x_\mu-y_\mu)}}{(2\pi)^3\,2\omega(\vec{k})}.
\end{equation} Since the commutator function is linked to the Feynman propagator we will have the interpretation for the probability for the propagation of field quanta between the nuclear spins located at $x$ and $y$.

\section{Algebraic Quantum Field Theory}
Operator algebras are central objects in the algebraic reformulation of Quantum Statistical Mechanics and Quantum Field Theory. Several structural elements of operator algebras are required for the numerical calculations in the application of mQED at finite temperatures to NMR. Therefore, some mathematical basics of operator algebras are briefly reviewed from \cite{BratteliRobinsonI} and \cite{BratteliRobinsonII} bevor the Field Theory is described.  
\medskip \\
\textbf{Basics of Operator Algebras.} 
The commutant of an algebra $\mathfrak{A}$ is denoted by $\mathfrak{A}'$ and we have $(\mathfrak{A}')'=\mathfrak{A}''$. The set of bounded operators on a Hilbert space $\mathfrak{H}$ is denoted by $\mathfrak{B}(\mathfrak{H})$.
\medskip \\
\textbf{Definition 1:}
A \textit{von Neumann algebra} on a Hilbert space $\mathfrak{H}$ is a $^*$-subalgebra $\mathfrak{M}$ of $\mathfrak{B}(\mathfrak{H})$ such that
\begin{equation}
\mathfrak{M}=\mathfrak{M}''.
\end{equation}
The terminology $W^*$-algebra is often used for the abstractly defined algebra and then the name von Neumann algebra is reserved for the operator algebras. 
 Note that a $C^*$-algebra is a closed set in the norm topology and a $W^*$-algebra is weakly closed. A bounded observable $A$ is a selfadjoint element of a
$C^*$- or a $W^*$-algebra $\mathfrak{A}$.
A $^*$-morphism $\pi$ between two $^*$-algebras $\mathfrak{C}$ and $\mathfrak{B}$ is
defined as a mapping $\pi:A\in\mathfrak{C}\longrightarrow\pi(A)\in\mathfrak{B}$
for all $A\in\mathfrak{C}$ such that $\pi(\alpha A+\gamma C)=\alpha\pi(A)+\gamma\pi(C)$, $\pi(AC)=\pi(A)\pi(C)$, and $\pi(A^*)=\pi(A)^*$
 for all $A,C\in\mathfrak{C}$ and $\alpha,\gamma\in\mathbb{C}$.
The kernel of a $^*$-morphism is given by the set $ker(\pi)=\{A\in\mathfrak{A};\pi(A)=0\}$. A state $\omega$ is a positive, normalized, and linear functional on $\mathfrak{A}$, i.e., $\omega\in\mathfrak{A}^*$, where $\mathfrak{A}^*$ is the dual of $\mathfrak{A}$. An expectation value is given by
$\omega(A)=(\psi_\omega,\pi_\omega(A)\psi_\omega)$, where 
  $\pi_\omega:\mathfrak{A}\rightarrow\mathfrak{B}(\mathfrak{H})$ and
$\psi_\omega\in\mathfrak{H}_\omega$, where the index $\omega$ denotes the
association of the representation 
$(\mathcal{H}_\omega,\pi_\omega)$ with the state $\omega$. The space $\mathfrak{H}_\omega$ is called the representation space and the operator
examples $\pi(A)$ are called the representatives of $\mathfrak{A}$. The representation is said to be faithful if, and only if, $\pi_\omega$ is a
$^*$-isomorphism between
$\mathfrak{A}$ and $\pi(\mathfrak{A})$, i.e., if, and only if, $ker(\pi_\omega)=\{0\}$.
A faithful representation satisfies $\parallel\pi_\omega(A)\parallel=\parallel
A\parallel$, for all $A\in\mathfrak{A}$.  If $(\mathfrak{H},\pi)$ is a representation of the $C^*$-algebra $\mathfrak{A}$
and if $\mathfrak{H}_0$ is a subspace of $\mathfrak{H}$, then $\mathfrak{H}_0$ is
said to be invariant under $\pi$ if $\pi(A)\mathfrak{H}_0\subseteq\mathfrak{H}_0$ for all
$A\in\mathfrak{A}$. Hence, if $\mathfrak{H}_0$ is invariant under $\pi$ and $\mathfrak{H}^\perp$ is the orthogonal
complement of $\mathfrak{H}_0$, i.e., $\mathfrak{H}^\perp\dot{=}\{\xi\in\mathfrak{H};\langle
\xi,\psi\rangle=0,\,\forall\psi\in\mathfrak{H}_0\}$, then we have $\langle \xi,\pi(A)\psi\rangle=0$ for all $A\in\mathfrak{A}$ and all $\xi\in\mathfrak{H}^\perp$,
$\psi\in\mathfrak{H}_0$. A $^*$-isomorphism of an algebra $\mathfrak{A}$ into itself is called a
$^*$-automorphism $\tau$. The time evolution of a physical system is given by a one-parametric group of $^*$-automorphisms $\tau_t$, which is generated by a derivation $\delta$. Thus, the derivation $\delta$ contains the information of the Hamiltonian $H$ and one formally has $A\mapsto \tau_t(A)=e^{\frac{t}{\hbar}\delta}(A)$. 
\\ \medskip \\
\textbf{Definition 2:}
A $W^*$-dynamical system is a pair $(\mathfrak{M},\tau)$, where $\mathfrak{M}$ is a $W^*$-algebra and $\tau: G\rightarrow \mathrm{Aut(\mathfrak{M})}$, $G\ni g\mapsto \tau_g$ is a weakly continuous representation of a locally compact group $G$ as *-automorphisms acting on $\mathfrak{M}$.
\\ \medskip \\
Note that a $C^*$-dynamical system $(\mathfrak{A},\tau)$ is defined in a similar fashion. In this case $\mathfrak{A}$ is a $C^*$-algebra and $\tau$ is a strongly continuous representation of a locally compact group as *-automorphisms acting on $\mathfrak{A}$. In order to proceed with equilibrium states we define the strip $S_\beta\dot{=}\{z\in\mathbb{C}\mid 0 <\Im (z)<\beta \}$.
\\ \medskip \\
\textbf{Definition 3:} Let $(\mathfrak{A},\tau)$ be a $C^*$- or a $W^*$-dynamical system. A state $\omega^\beta$ on $\mathfrak{A}$, supposed to be normal in the $W^*$-case, is a $(\tau,\beta)$-KMS state for some $\beta\in\mathbb{R}^+$ if the following holds. For any $A,B\in\mathfrak{A}$ there exist a function $F_\beta(A,B;z)$ which is analytic on the strip $S_\beta$, continuous on its closure and satisfies the Kubo-Martin-Schwinger condition 
\begin{equation}
 F_\beta(A, B; t)=\omega^\beta(A\tau_t(B))\quad \mathrm{and} \quad F_\beta(A, B; t+i\beta)= \omega^\beta(\tau_t(B)A)
\end{equation}
on the boundary of $S_\beta$.
\\ \medskip \\
\textbf{Description of the field theory}. A single boson is described as a square integrable function $f\in\mathfrak{H}=L^2(\mathbb{R}^3)$ in position or momentum space $\mathbb{R}^3$. The Hilbert space $\mathfrak{H}$ is called the 1-particle Hilbert space. The n-particle Hilbert space $\mathfrak{H}^n$ is given by the n-fold tensor product of $\mathfrak{H}$ with itself, i.e.,
$\mathfrak{H}^n=\mathfrak{H}\otimes\mathfrak{H}\cdots\otimes\mathfrak{H}$. The projection $P_+\mathfrak{H}^n=\mathfrak{H}^n_+$ \cite{BratteliRobinsonII} onto totally symmetric n-particle wave functions reflects that the particles obey the Bose-Einstein statistics. The bosonic Fock-space is then defined by
\begin{equation}
\mathfrak{F}_+(\mathfrak{H})=\bigoplus^\infty_{n=0}\mathfrak{H}^n_+.
\end{equation}  We have $\mathfrak{H}^0=\mathbb{C}$ and the vacuum is described by $\Omega_0=(1,0,0,...)\in\mathfrak{F}_+(\mathfrak{H})$. The smeared creation and annihilation operators are defined by
\begin{equation}
a^*_\lambda(f)=\int d^3k\, f(\vec{k})a_\lambda^*(\vec{k}) \quad \mathrm{and}\quad   a_\lambda(f)=\int d^3k\, \overline{f}(\vec{k})a_\lambda^*(\vec{k})  
\end{equation} for $f\in \mathfrak{H}$. $a_{\lambda}^*(\vec{k})$ and $a_{\lambda}(\vec{k})$ satisfy the commutation relations in eq. (\ref{eq:commutation}), which translates to 
\begin{equation}
[a_{\lambda}(f),a_{\lambda'}^*(g)]=\delta_{\lambda,\lambda'} \langle f\mid g\rangle_\mathfrak{H}\quad \mathrm{and} \quad [a_{\lambda}(f),a_{\lambda'}(g)]= [a_{\lambda}^*(f),a_{\lambda'}^*(g)]= 0
\end{equation}  with scalar product $\langle\cdot\mid\cdot\rangle$ on $\mathfrak{H}$ given by
\begin{equation}
\langle f\mid g\rangle_\mathfrak{H}=\int_{\mathbb{R}^3}d^3 k\, \bar{f}(k)g(k).
\end{equation} The notation for the quantized magnetic field in section II is recovered by
\begin{equation}
B^\alpha_\varphi(x)\equiv\Phi(b^{\alpha }_{\varphi}(x))\dot{=} \frac{1}{\sqrt{2}}\sum_{\lambda=1,2}\big(a^{*}_{\lambda}(b^{\alpha x}_{\varphi\lambda})+a_{\lambda}(b^{\alpha x}_{\varphi\lambda})\big). 
\end{equation} According to eq. (\ref{eq:MagneticField}) the functions $b^{\alpha x}_{\varphi\lambda}:\mathbb{R}^3\rightarrow\mathbb{C}$ are given by
\begin{equation}
b^{\alpha x}_{\varphi\lambda}(k)= i\sqrt{\frac{\hbar}{\epsilon_0(2\pi)^3}}\big(\vec{k}\times\vec{\epsilon}_\lambda(\vec{k})\big)^\alpha\frac{\varphi(\vec{k})}{\sqrt{\omega(\vec{k})}}e^{-ik^\mu x_\mu} \quad\mathrm{and}\quad \alpha=\mathrm{x,y,z}.
\end{equation} 
Since the field operators are unbounded one introduces the bounded Weyl operators
\begin{equation}
 W(f)=\mathrm{exp}(i\Phi(f)),\quad\mathrm{satisfying}\quad W(f)W(g)=e^{-i\Im (\langle f\mid g\rangle_{\mathfrak{H}})}W(f+g). 
\end{equation} In order to rigorously define equilibrium states the one-particle Hilbert space has to be restricted by
$\mathfrak{H}^\mathrm{r}=\{ f\in\mathfrak{H}; \omega^{-1/2}f\in\mathfrak{H}\}$, which ensures a suitable infrared behavior. This basically means to "reduce or neglect" extremely low energetic photons. However, in this document no infrared divergences were found in the numerical calculations and the restriction of $b^{\alpha x}_{\varphi\lambda}$  to $\mathfrak{H}^\mathrm{r}$ can be chosen such that the influence of the restriction on the expectation value is arbitrarily small. We define a $C^*$-algebra $\mathfrak{A}_{\mathrm{EM}}$ for the quantized electromagnetic field by
\begin{equation}\label{eq:AlgebraEM}
\mathfrak{A}_{\mathrm{EM}}\dot{=}\mathcal{W}(\mathfrak{H}^\mathrm{r})=\overline{\mathrm{span}\{W(f);\,f\in\mathfrak{H}^\mathrm{r} \}}^{\parallel \cdot \parallel},
\end{equation} where the closure is taken in the uniform norm $\parallel \cdot \parallel$ for bounded operators on the bosonic Fock space $\mathfrak{F}_+(\mathfrak{H}^\mathrm{r})$. The dispersion relation is given by $\omega(\vec{k})=c\mid \vec{k}\mid$ where $c$ is the speed of light and the free field Hamiltonian is given by
\begin{equation}
H_\mathrm{EM}\,\dot{=}\,d\Gamma(\omega)\,\equiv\,\hbar\int_{\mathbb{R}^3}d^3 k\,\omega(\vec{k})a^*(\vec{k})a(\vec{k}).
\end{equation} $d\Gamma(\omega)$ provides an infinitesimal generator $\delta_\mathrm{EM}$, formally given by $\delta_\mathrm{EM}=[H_\mathrm{EM},\cdot]$, that generates the one-parameter group $\{\tau^\mathrm{EM}_t \}_{t\in\mathbb{R}}$ for the quantized electromagnetic field. This group provides the free field dynamics and the action is given by 
\begin{equation}
W(f)\mapsto\tau^\mathrm{EM}_t(W(f))=W(e^{i\omega t} f)  \quad \mathrm{which\,implies}\quad \Phi(f)\mapsto\tau^\mathrm{EM}_t(\Phi(f))=\Phi(e^{i\omega t} f).
\end{equation} This is also known as Bogoliubov transformation. Note that the group $\{\tau^\mathrm{EM}_t\mid t\in\mathbb{R}\}$ is not strongly continuous because $\parallel W(f)-W(g) \parallel=2$ $\forall g\neq f$ and hence $(\mathcal{W}(\mathfrak{H}^\mathrm{r}),\tau^\mathrm{EM})$ is not a $C^*$-dynamical system. 
\medskip \\
The GNS-representation $(\mathfrak{H}_\mathrm{AW},\pi_\mathrm{AW}^\beta)$ which is induced by the $(\tau^{\mathrm{EM}},\beta)$-KMS state $\omega^\beta_{\mathrm{EM}}$ on $\mathfrak{A}_{\mathrm{EM}}$ was found by Araki and Woods \cite{araki1963representations} and is therefore referred as Araki-Woods representation. The representation space is given by
\begin{equation}
\mathfrak{H}_\mathrm{AW}=\mathfrak{F}_+(\mathfrak{H}^\mathrm{r})\otimes\mathfrak{F}_+(\mathfrak{H}^\mathrm{r})\otimes\mathfrak{F}_+(\mathfrak{H}^\mathrm{r})\otimes\mathfrak{F}_+(\mathfrak{H}^\mathrm{r}), 
\end{equation} and the annihilation operators are given by
\begin{equation}
\pi_\mathrm{AW}^\beta(a_{1}(f))=\big( a^*\big(\sqrt{1+\rho_\beta}f\big) \otimes \hat{1}\otimes \hat{1}\otimes\hat{1}+\hat{1}\otimes a\big(\sqrt{\rho_\beta}\bar{f}\big)\big)\otimes \hat{1}\otimes\hat{1}
\end{equation} and 
\begin{equation}
\pi_\mathrm{AW}^\beta(a_{2}(f))=\hat{1}\otimes\hat{1}\otimes \big( a^*\big(\sqrt{1+\rho_\beta}f\big) \otimes \hat{1}+\hat{1}\otimes \hat{1}\otimes\hat{1}\otimes a\big(\sqrt{\rho_\beta}\bar{f}\big)\big).
\end{equation} The function $\rho_\beta$ is a physical input which ensures that Planck's law for the thermal radiation density and Bose-Einstein statistics is satisfied and we have 
\begin{equation}
\rho_\beta(\vec{k})=\frac{1}{e^{\beta \omega(\vec{k})}-1}.  
\end{equation} The vector representative $\Omega^\beta_\mathrm{AW}$ of $\omega^\beta_{\mathrm{EM}}$ is cyclic and separating for the weak closure 
$\pi_\mathrm{AW}^\beta\big(\mathfrak{A}_{\mathrm{EM}} \big)''$ of $\mathfrak{A}_{\mathrm{EM}}$ and it turns out that $\big(\pi_\mathrm{AW}^\beta\big(\mathfrak{A}_{\mathrm{EM}} \big)'',\{\pi_\mathrm{AW}^\beta\circ\tau^\mathrm{EM}_t \}_{t\in\mathbb{R}}\big)$ is a $W^*$-dynamical system \cite{muck2005thermal}. Using $(\mathfrak{H}_\mathrm{AW},\pi_\mathrm{AW}^\beta)$ it can be derived that
\begin{equation}
\omega^\beta_{\mathrm{EM}}(\tau^\mathrm{EM}_{z_2}\big(\hat{\Phi}( b^{\alpha }_{\varphi}(\vec{x})))\big)\tau^\mathrm{EM}_{z_1}\big(\hat{\Phi}(b^{\gamma }_{\varphi}(\vec{y}))\big)=\sum_{\lambda=1,2}\int_{\mathbb{R}^3}d^3k\bigg( \,\overline{b^{\alpha(\vec{x},z_2)}_{\varphi\lambda}(\vec{k})} b^{\gamma(\vec{y},z_1)}_{\varphi\lambda}(\vec{k})(1+\rho_\beta(\vec{k}))+\overline{b^{\gamma(\vec{y},z_1)}_{\varphi\lambda}(\vec{k})}b^{\alpha(\vec{x},z_2)}_{\varphi\lambda}(\vec{k})\rho_\beta(\vec{k}) \bigg).
\end{equation} For later purpose we define the \textit{magnetic quantum exchange} $\mathfrak{m}^{\alpha\gamma}_{\varphi\beta}:\mathbb{R}^3\times Z\times\mathbb{R}^3\times Z\times\mathbb{R}^3\rightarrow\mathbb{C}$ with the strip $Z=[0,\infty)\times[0,i\beta)$ in the complex plane $\mathbb{C}$ by
\begin{align}
(\vec{x},z_2,\vec{y},z_1,\vec{k})\mapsto\mathfrak{m}^{\alpha\gamma}_{\varphi\beta}(\vec{x},z_2,\vec{y},z_1,\vec{k})\dot{=}\sum_{\lambda=1,2}\bigg( &\overline{b^{\alpha(\vec{x},z_2)}_{\varphi\lambda}(\vec{k})} b^{\gamma(\vec{y},z_1)}_{\varphi\lambda}(\vec{k})(1+\rho_\beta(\vec{k}))+
\overline{b^{\gamma(\vec{y},z_1)}_{\varphi\lambda}(\vec{k})} b^{\alpha(\vec{x},z_2)}_{\varphi\lambda}(\vec{k})\rho_\beta(\vec{k})
\\ \nonumber+&\overline{b^{\gamma(\vec{y},z_2)}_{\varphi\lambda}(\vec{k})} b^{\alpha(\vec{x},z_1)}_{\varphi\lambda}(\vec{k})(1+\rho_\beta(\vec{k}))+
\overline{b^{\alpha(\vec{x},z_1)}_{\varphi\lambda}(\vec{k})} b^{\gamma(\vec{y},z_2)}_{\varphi\lambda}(\vec{k})\rho_\beta(\vec{k})\bigg).
\end{align}
The following useful symmetry is valid:  $\mathfrak{m}^{\alpha\gamma}_{\varphi\beta}(\vec{x},z_2,\vec{y},z_1,\vec{k})=\mathfrak{m}^{\alpha\gamma}_{\varphi\beta}(\vec{x},z_1,\vec{y},z_2,\vec{k})$. In applications to NMR it turns out that the family $\{\mathfrak{m}^{\alpha\gamma}_{\varphi\beta}\}_{\alpha,\gamma=\mathrm{x,y,z}}$ takes a central role for the strength and occurrence of the magnetic shielding (chemical shift) and determines return to equilibrium properties. 

\section{Quantum Spin Systems and Spin Boson Systems}\label{sectionX}
In the perturbation series used in this document Quantum Spin Systems (QSS) occur as subsystems of Spin Boson Systems (SBS) while SBS occur as subsystems of the mQED systems. 
\medskip \\
\textbf{Quantum Spin Systems.}
The mathematical framework for QSS is taken from \cite{BratteliRobinsonII, ThemIV}.
A quantum spin system consists of particles on a lattice $\mathbb{Z}^d$. We associate with each point $x\in\mathbb{Z}^d$ a
Hilbert space $\mathfrak{H}_{x}$ of dimension $2s(x)+1$ and with a finite subset $\lambda=\{x_1,...,x_v\}\subset\mathbb{Z}^d$ we associate the
tensor product space $\mathfrak{H}_\Lambda=\bigotimes_{x_i\in\Lambda}\mathfrak{H}_{x_i}$. 
The lattice can be equibed with a metric $d(\cdot,\cdot)$.
The local physical observables are contained in the algebra of all bounded operators acting on $\mathfrak{H}_\Lambda$, that is the local 
$C^*$-algebra $\mathfrak{A}_\Lambda\cong\bigotimes_{x_i\in\Lambda}M_{2s(x_i)+1}$ in which $M_n$ denote the algebra of $n\times n$ complex matrices. 
If $\Lambda_1\cap\Lambda_2=\emptyset$, then $\mathfrak{H}_{\Lambda_1\cup\Lambda_2}=\mathfrak{H}_{\Lambda_1}\otimes\mathfrak{H}_{\Lambda_2}$ and  $\mathfrak{A}_{\Lambda_1}$
is isomorphic to the 
$C^*$-subalgebra $\mathfrak{A}_{\Lambda_1}\otimes\hat{1}_{\Lambda_2}$ of $\mathfrak{A}_{\Lambda_1\cup\Lambda_2}$, where $\hat{1}_{\Lambda_2}$ denotes the identity 
operator on $\mathfrak{H}_{\Lambda_2}$. If $\Lambda_1\subseteq\Lambda_2$ then
$\mathfrak{A}_{\Lambda_1}\subseteq\mathfrak{A}_{\Lambda_2}$ and  operators with disjoint support commute, i.e.
$[\mathfrak{A}_{\Lambda_1},\mathfrak{A}_{\Lambda_2}]=0$ whenever $\Lambda_1\cap\Lambda_2=\emptyset$.
We may define the algebra of "all local observables" as $\mathfrak{A}_{loc}=\bigcup_{\Lambda\subset\mathbb{Z}^d}\mathfrak{A}_\Lambda$. The operator norm of an element $A\in\mathfrak{A}_\Lambda$ is given by 
$\parallel A\parallel=\sup\{\parallel A\Psi\parallel;\Psi\in\mathfrak{H}_\Lambda,\parallel \Psi\parallel=1\}$. 
An $\epsilon$-neighborhood of an operator $A$ is the set of operators $B$ with $\parallel A-B\parallel\leq\epsilon$ \cite{ThemIII}.  
The local convex topology which is induced by the operator norm is called the uniform topology and 
the quantum spin algebra $\mathfrak{A}$ is then obtained by taking the closure of the algebra $\mathfrak{A}_{loc}$ in this topology, i.e. 
$\mathfrak{A}=\overline{\mathfrak{A}_{loc}}^{\parallel\cdot\parallel}$. An interaction $\Phi$ is defined to be a function from a finite subset $X\subset\mathbb{Z}^d$ into the hermitian elements of $\mathfrak{A}$ such that
$\Phi(X)\in\mathfrak{A}_X$. The Hamiltonian associated with the region $\Lambda$ is then given by
\begin{equation}
 H_\Phi(\Lambda)=\sum_{X\subseteq\Lambda}\Phi(X).
\end{equation} An interaction of a spin with a classical, external magnetic field \cite{Them, ThemII, ThemFokus} is given by
\begin{equation}
\Phi(\{j\})= \gamma_j\vec{I}_j \cdot\vec{B}_\mathrm{ext} \quad \mathrm{for \,nuclear\, spins\, and} \quad \Phi(\{i\})=g_\mathrm{S}\frac{\mu_\mathrm{B}}{\hbar}\vec{S}_i\cdot\vec{B}_\mathrm{ext} 
\quad  \mathrm{for\, spins\, of \,electrons.}
\end{equation} An NMR pulse induces a time-dependent interaction $P_t$ involving spins and oscillating, external magnetic fields \cite{BG}, \cite{ThemMSC} 
\begin{equation}\label{eq:Pulse}
P_t\,\dot{=}\,\sum_{j=1}^K \Phi^\mathrm{P}_t(\{j\}), \quad \mathrm{where}\quad  \Phi^\mathrm{P}_t(\{j\})= \gamma_j\vec{I}_j \cdot\vec{B}_\mathrm{ext}(t).
\end{equation} For example, a single pulse in x-direction, which is switched on from time $t=0$ to $t=t_0$, is described by a magnetic field of the form
\begin{equation}
B^\mathrm{y}_\mathrm{ext}(t)=B_\mathrm{P}\int d\omega_\mathrm{P} f(\omega_\mathrm{P})\theta(t,0,t_0)\cos(\omega_\mathrm{P}t+\phi).
\end{equation} $\theta$ is the step function, $B_\mathrm{P}$ provides the amplitude of the pulse (some milli Tesla), $\phi$ is the phase of the magnetic field at $t=0$ and $f$ provides the frequency distribution of the pulse. Often, the frequency distribution provided by $f$ is of rectangular form and of course it has to cover the excitation frequencies of the nuclei which shall be excited.  
The dynamical evolution of an observable $A\in\mathfrak{A}_\Lambda$ for a system with time-independent Hamiltonian $H_\Phi(\Lambda)\in\mathfrak{A}_\Lambda$ can be described by the Heisenberg relations
\begin{equation}
 \tau^{\mathrm{S}\Lambda}_t:\mathfrak{A}_\Lambda\rightarrow \mathfrak{A}_\Lambda , \quad A\mapsto \tau^{\mathrm{S}\Lambda}_t(A)=e^{\frac{itH_\Phi(\Lambda)}{\hbar}}Ae^{-\frac{itH_\Phi(\Lambda)}{\hbar}}.
\end{equation} Thus the map $t\in\mathbb{R}\mapsto\tau^{\mathrm{S}\Lambda}_t$ is a one-parameter group of $^*$-automorphisms of the matrix 
algebra $\mathfrak{A}_\Lambda$ and S denotes that this automorphism group acts only on the quantum spin algebra. The corresponding derivation is denoted by $\delta_\Lambda$ and $(\mathfrak{A}_\Lambda,\tau^{\mathrm{S}\Lambda}_t)$ is a $C^*$-dynamical system because $\tau^{\mathrm{S}\Lambda}_t$ is strongly continuous for finite external fields. Since effective spin-spin couplings are absent in the mQED Hamiltonian eq. (\ref{eq:FreeQED}) and (\ref{eq:IntQED}) a spin system consisting of $K$ nuclei and $E$ electrons forms a subsystem of (\ref{eq:FreeQED}) whose equilibrium state is given by
\begin{equation}
\omega^\beta_\mathrm{S}=\bigotimes_{j=1}^{K+E}\omega^\beta_{\mathrm{S}j}.
\end{equation} $\omega^\beta_{\mathrm{S}j}$ is the $(\tau^{\mathrm{S}j},\beta)$-KMS state of the single nucleus or electron enumerated by $j$. The representation which is induced by $\omega^\beta_\mathrm{S}$ is denoted by $(\mathfrak{H}_\mathrm{S},\pi_\mathrm{S})$.
\medskip \\
\textbf{Perturbative description of Spin Boson Systems.}
A $C^*$-algebra $\mathfrak{A}_\mathrm{SB}$ for spins located in $\Lambda$ interacting with bosons from the quantized electromagnetic field is given by
\begin{equation}
\mathfrak{A}_\mathrm{SB}\dot{=}\overline{\mathrm{span}\{A\otimes W(f)\mid A\in\mathfrak{A}_\Lambda,f\in\mathfrak{H}^\mathrm{r}}\}^{\parallel\cdot\parallel_{\mathfrak{B}(\mathfrak{H}_{\mathrm{SB}})}}=\mathfrak{A}_\Lambda\otimes\mathfrak{A}_{\mathrm{EM}},
\end{equation} where $\mathfrak{H}_{\mathrm{SB}}=\mathfrak{H}_\Lambda\otimes\mathfrak{H}^\mathrm{r}_{\mathrm{AW}}$ is a representation space. The index $\Lambda$ is neglected for simplicity.
The free time evolution $\tau^{\mathrm{SB}}_t=\tau^{\mathrm{S}\Lambda}_t\otimes\tau^{\mathrm{EM}}_t$, with derivation $\delta_\mathrm{SB}=\delta_\Lambda+\delta_\mathrm{EM}$, acts on $\mathfrak{A}_\mathrm{SB}$ and we have $\tau^{\mathrm{SB}}_t(A\otimes W(f))\in\mathfrak{A}_\mathrm{SB}$ \cite{muck2005thermal}. Interactions of the form 
\begin{equation}\label{eq:SBSInt}
H^\mathrm{SB}_\mathrm{Int}=\sum_{\alpha=\mathrm{x,y,z}}\bigg(\sum_{j=1}^K\gamma_j I^\alpha_j\otimes\Phi(b^{\alpha }_{\varphi}(\vec{x}_j))+
\sum_{i=1}^E g_\mathrm{S}\frac{\mu_\mathrm{B}}{\hbar}S^\alpha_i\otimes\Phi(b^{\alpha }_{\varphi}(\vec{x^\mathrm{e}_i}))\bigg)
\end{equation} enable energy exchange between spins and bosons. Interactions given by (\ref{eq:SBSInt}) are unbounded and if the derivation induced by $H^\mathrm{SB}_\mathrm{Int}$ is denoted by $\delta^\mathrm{Int}_\mathrm{SB}$ then the
evolution group $\{\tau^{I\mathrm{SB}}_t\}_{t\in\mathbb{R}}$ generated by $\delta^I_\mathrm{SB}=\delta^\mathrm{f}_\mathrm{SB}+\delta^\mathrm{Int}_\mathrm{SB}$ does not necessarily leaves $\mathfrak{A}_\mathrm{SB}$ invariant. However, if some general conditions are satisfied \cite{muck2005thermal} $\tau^{I\mathrm{SB}}_t(A)$ lies in the weak closure $\mathfrak{A}_\mathrm{SB}''$, i.e. $\tau^{I\mathrm{SB}}_t(\mathfrak{A}_\mathrm{SB})\subseteq \mathfrak{A}_\mathrm{SB}''$. Furthermore, if the conditions from \cite{muck2005thermal} are satisfied the convergence of the right hand side of
\begin{equation}\label{eq:SBDynamics}
\tau^{I\mathrm{SB}}_t(A)=\tau^{\mathrm{SB}}_t(A)+\sum_{n\geq1}i^n\int^t_0dt_1\int^{t_1}_0dt_2\cdots\int^{
t_{n-1}}_0dt_n[\tau^{\mathrm{SB}}_{t_n}(H^\mathrm{SB}_\mathrm{Int}),[\cdots[\tau^{\mathrm{SB}}_{t_1}(H^\mathrm{SB}_\mathrm{Int}),\tau^{\mathrm{SB}}_{t}(A)]]]
\end{equation} towards $\tau^{I\mathrm{SB}}_t(A)$ holds strongly on vectors of the form $\mid \Omega\rangle=\mid \Omega_\Lambda\rangle\otimes\mid \Omega_0\rangle$ and observables of the form $A=I\otimes W(f)$, where $I\in\mathfrak{A}_\Lambda$ and $\mid \Omega_\Lambda\rangle\in\mathfrak{H}_\Lambda$. For a large class of coupling functions $\varphi$ \cite{muck2005thermal} the pair 
\begin{equation}
\big(\pi_\mathrm{SB}^\beta(\mathfrak{A}_\mathrm{SB})'',\pi_\mathrm{SB}^\beta\circ\tau^{I\mathrm{SB}}\big)
\end{equation} is a $W^*$-dynamical system and $\pi_\mathrm{SB}^\beta=\pi_\mathrm{S}\otimes\pi_\mathrm{AW}^\beta$.  An important state $\hat{\omega}^{I\beta}_{\mathrm{SB}}$ on the von Neumann algebra $\pi_\mathrm{SB}^\beta(\mathfrak{A}_\mathrm{SB})''$ is given by \cite{BratteliRobinsonII}
\begin{equation}\label{eq:ESBS}
 \hat{\omega}^{I\beta}_{\mathrm{SB}}(A)=\hat{\omega}^{\beta}_{\mathrm{SB}}(A)+\sum_{n\geq1}(-1)^n\int^\beta_0ds_1\int^{s_1}_0ds_2\cdots\int^{s_{n-1}}_0ds_n\,
 \hat{\omega}^{\beta}_{\mathrm{T,SB}}(A,\hat{\tau}^{\beta\mathrm{SB}}_{is_n}(H^{\mathrm{SB}}_\mathrm{Int}),...,\hat{\tau}^{\beta\mathrm{SB}}_{is_1}(H^\mathrm{SB}_\mathrm{Int})),
\end{equation} where $\hat{\omega}^{\beta}_{\mathrm{SB}}$ is the extension of $\omega^{\beta}_{\mathrm{SB}}=\omega^\beta_\mathrm{S}\otimes\omega^\beta_\mathrm{EM}$ on $\mathfrak{A}_\mathrm{SB}$ to 
$\pi_\mathrm{SB}^\beta(\mathfrak{A}_\mathrm{SB})''$, $\hat{\tau}^{\beta\mathrm{SB}}\dot{=}\pi_\mathrm{SB}^\beta\circ\tau^{\mathrm{SB}}$, $A\in\pi_\mathrm{SB}^\beta(\mathfrak{A}_\mathrm{SB})''$ and T denotes that truncated functions are used  \cite{BratteliRobinsonII}. If the conditions from 
\cite{muck2005thermal} are satisfied one finds for a large class of states $\eta$ and observables $A$ return to equilibrium for the interacting system, formally given by 
\begin{equation}
\lim_{t\rightarrow\infty}\eta\circ\tau^{I\mathrm{SB}}_t(A)= \omega^{I\beta}_{\mathrm{SB}}(A).
\end{equation} In this case $\omega^{I\beta}_{\mathrm{SB}}$ is a $(\tau^{I\mathrm{SB}},\beta)$-KMS state. For applications to NMR we define the evolution group $\{\tau^{I\mathrm{SB}}_{Pt}\}_{t\in\mathbb{R}}$ which is generated by $\delta^I_\mathrm{SB}+\delta_{Pt}$. For $A\in\mathfrak{A}_\Lambda$ we have $\delta_{Pt}(A)=i[P_t,A ]$ and 
$t\in\mathbb{R}\mapsto P_t=P^*_t\in\mathfrak{A}_\Lambda$ is a one-parameter family of selfadjoint elements which contains the information of the pulse sequence given by eq. (\ref{eq:Pulse}). From now on we make the identification $\mathfrak{M}=\mathfrak{A}_\mathrm{SB}''$. Although there exist not yet a rigorous proof it seems to be obvious \cite{Fredenhagen} that if $\big(\pi_\mathrm{SB}^\beta(\mathfrak{A}_\mathrm{SB})'',\pi_\mathrm{SB}^\beta\circ\tau^{I\mathrm{SB}}\big)$ is a $W^*$-dynamical system then $\big(\pi_\mathrm{SB}^\beta(\mathfrak{M}),\pi_\mathrm{SB}^\beta\circ\tau^{I\mathrm{SB}}_P\big)$ is a $W^*$-dynamical system for a suitable class of pulse sequences $P$. 

\section{Application to NMR}
A typical NMR experiment consists basically of molecules interacting with external magnetic fields. In most experiments the interacting system is in thermal equilibrium at the beginning of the experiment. The molecular structures are then investigated by the application of a pulse sequence which consists of oscillating, external magnetic fields. Pulse sequences provide an out of equilibrium nuclear spin dynamics and they act only for a short time at the beginning of the experiment. When the pulse sequence is finished the system is again governed by the equilibrium dynamics which is then responsible for a return to equilibrium. This equilibration process is experimentally detected in NMR and referred as \textit{free induction decay} (FID). In most experiments the x- and y-components of the nuclear spins are recorded, while the z-component is not recorded. The detected FID is called NMR signal, $\langle M^+ \rangle(t)$, and its Fourier transform provides the NMR spectrum $S(\nu)$. An NMR spectrometer detects the radiation from the magnetically excited nuclear spins which is identical to the time evolution of the x- and y-components of the nuclear spins. 
Therefore, the NMR spectrometer records the NMR signal $\langle M^+ \rangle(t)=\sum_j \langle I^+_j(t) \rangle$ which consists of expectation values with observables $I^+_j(t)=I^\mathrm{x}_j(t)+iI^\mathrm{y}_j(t)$. The real and imaginary parts of the NMR signal are given by $\Re(\langle M^+ \rangle(t))=\sum_j\langle I^\mathrm{x}_j(t) \rangle$ and 
$\Im(\langle M^+ \rangle(t))=\sum_j\langle I^\mathrm{y}_j(t) \rangle$ respectively. In certain cases, the NMR spectrum contains only very sharp peaks of "Lorentzian shape". NMR spectra which show any other distribution may be obtained by (continuous) superpositions of Lorentz functions. It is usually seen that the positions of the peaks are shifted towards lower frequencies compared to the \textit{Lamor frequency} $v_0=\gamma \mid B^{\mathrm{z}}_\mathrm{ext} \mid$. This is called \textit{chemical shift} and it is a direct consequence of the magnetic shielding which is caused by the electrons: In the presence of an external magnetic field the magnetic moments of electrons show into the opposite direction compared to the magnetic moments of the nuclei. Hence, the external magnetic field at the position of a nucleus is reduced (shielded) by electrons.    
\begin{figure}[h]     
\centering
\includegraphics*[scale=0.85]{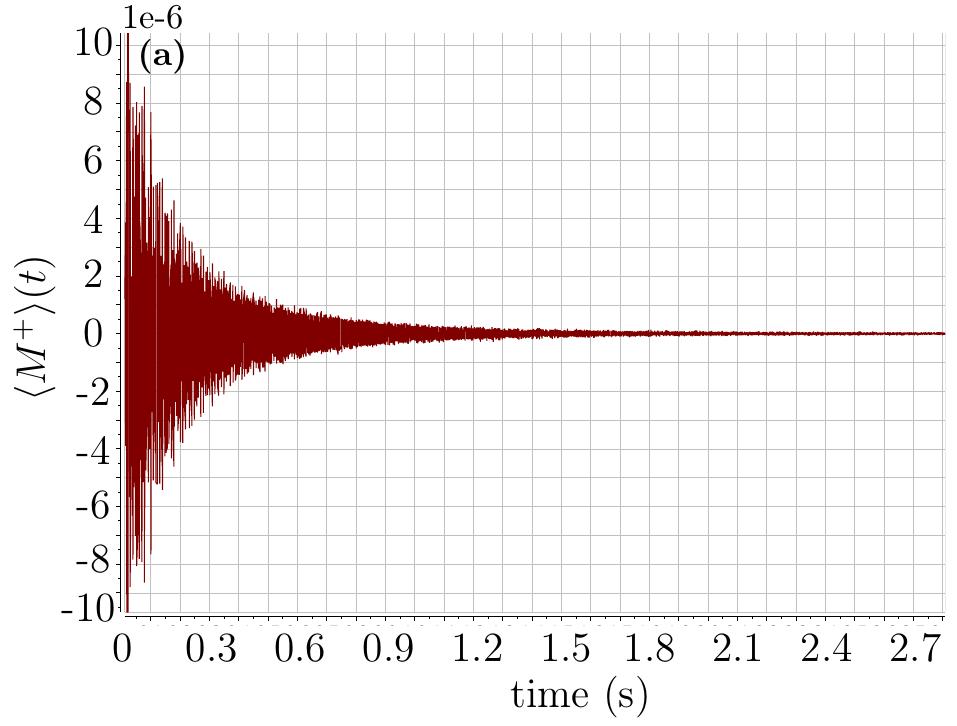}
\includegraphics*[scale=0.85]{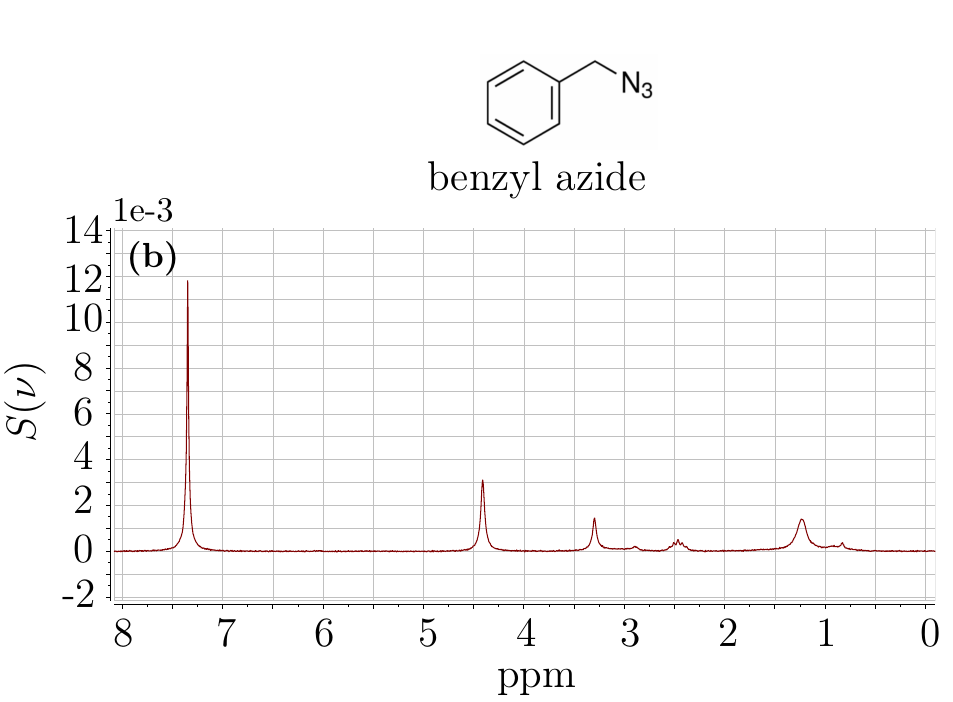}
\caption{\label{fig:ExpData} Example of an experimental 1H NMR signal $\langle M^+ \rangle(t)$ (left side) and the corresponding NMR spectrum $S(\nu)$ (right side). There is an exponentially fast return to equilibrium of $\langle M^+ \rangle(t)$ (a). The peaks in the spectrum (b) possess different line widths and originate from protons with different electronic environment. A discrete spectrum is unable to give an accurate description of the domain around 1 ppm. The presence of several peaks provides oscillations in the NMR-signal decay. A smooth exponential decay is obtained if only identical protons contribute to the signal, e. g. the protons from water. }
\end{figure}

Fig. (\ref{fig:ExpData}) shows exemplary 1H NMR-data of 12$\mu$l benzyl azide with impurities in dimethyl sulfoxide-D6. The NMR signal $\langle M^+ \rangle(t)$ is shown in (a) and the spectrum $S(\nu)$ in (b). An exponentially fast thermalization for $\langle M^+ \rangle(t)$ can be seen on the left image. Note that this is the thermalization for the expectation values of x- and y-components of nuclear spin operators. The z-component needs an equal or more time for return to equilibrium. The Fourier transform (right image) provides the frequencies involved in the nuclear spin dynamics during thermalization. It can be seen that the peaks possess different widths and positions. The domain around 1 and 3 ppm is generated only by a few protons but the description of this domain by a discrete spectrum is not accurate. Hence, model-calculations involving a continuous spectrum are desirable for a detailed analysis of this spectrum. The different line widths contain important information about the distribution of the electrons and nuclei.

According to the description of an experimental NMR setup the mQED system in the algebraic framework is applied as follows. At times before pulse sequences, $t\leq0$, the description of the molecular system interacting with the classical and quantized electromagnetic field will be described by the Hamiltonian $H=H_0+H^\mathrm{SB}_\mathrm{Int}$ from eq. (\ref{eq:FreeQED}) and (\ref{eq:SBSInt}). $H^\mathrm{SB}_\mathrm{Int}$ contains the same interactions as $H_\mathrm{Int}$ from eq. (\ref{eq:IntQED}) except the interactions which does not involve spins. Hence, it may be referred as Spin Boson approximation of mQED. This approximation is based on the assumption that the energy of a pulse sequence is too low to change the momentum and geometry of the investigated electronic structure. 
 This approximation is also made by the effective spin model and there seems to be no obvious reason why this approximation should be unsuitable. For $t=0$ the system is in thermal equilibrium and the equilibrium state $\mid\Omega^\beta\rangle$ is determined by $H$. Pulse sequences are initiated at $t=0$, such that for $t>0$ the system is described by $H+P_t$. The time-dependent operator $P_t$ from eq. (\ref{eq:Pulse}) contains the information of the pulse sequence. Hence, the time-dependence of a nuclear spin operator, e.g., $I^\mathrm{z}(t)$ or $I^+(t)$, during a pulse sequence is determined by $H+P_t$. 

 Let $\mathfrak{B}(\mathfrak{H}_\mathrm{M})$ denote the set of bounded operators on the Hilbert space $\mathfrak{H}_\mathrm{M}$ of the molecular system. The spatial structure of the molecular system is contained in the state $\omega^\beta_\mathrm{M}:\mathfrak{B}(\mathfrak{H}_\mathrm{M})\rightarrow\mathbb{C}$. In the application to NMR the Born-Oppenheimer approximation will be used and the wave function $\psi$ of the electrons will be approximated by the ground state. Ideally the KMS state is used for the nuclear wave function $\Psi^\beta$. However, in most cases it is practically not possible to estimate this KMS state explicitly. This is because the nuclear Schr\"{o}dinger equation can only be solved for very simple molecules. Therefore, a suitable procedure may be used which approximates the square $\mid\Psi^\beta(X)\mid^2$ of the KMS state. A suitable initial choice may be given by inserting the potential energy surface (PSE) into the classical Gibbs state at inverse temperature $\beta$. If a  $\mid\Psi^\beta(X)\mid^2$ is found which shows agreement between experimental and calculated NMR spectra then a suitable approximation for the square of the KMS state for the nuclei is obtained. This may serve for molecular structure determination. For a function $f\in\mathfrak{B}(\mathfrak{H}_\mathrm{M})$ the expectation value is given by
\begin{equation}\label{eq:MolecularSystem}
\omega^\beta_\mathrm{M}(f)=\int d^{3K}x\int d^{3E}x^\mathrm{e}\mid \Psi^\beta(X) \mid^2 \mid \psi(X,X^\mathrm{e}) \mid^2 f(X,X^\mathrm{e}).
\end{equation} Remember that the dependence of $\tau_{Pt}^{I\mathrm{SB}}$ and $\hat{\omega}^{I\beta}_{\mathrm{SB}}$ on the coordinates $(X,X^\mathrm{e})$ was so far neglected in the notation for simplicity. For the main result this dependence is now written explicitly for clarity. Usually the temperature dependence is not explicitly indicated for the NMR-signal but we will do this in the following.
\medskip \\
\textbf{Main Result (NMR-signal from mQED for the reconstruction of $\mid \Psi^\beta(X) \mid^2$):}\\ \textit{Assume that} $\big(\mathfrak{M},\tau^{I\mathrm{SB}}_P\big)$ \textit{is a $W^*$-dynamical system and that $\omega^{I\beta}_{\mathrm{SB}}$ is a $(\tau^{I\mathrm{SB}},\beta)$-KMS state. Furthermore, let}
\begin{equation}
\tau_{Pt}^{I\mathrm{SB}}: \mathfrak{M}\rightarrow\mathfrak{M}, \quad \pi_\mathrm{SB}^\beta: \mathfrak{M}\rightarrow \mathfrak{B}(\mathfrak{H}_\mathrm{SB}) \quad and\quad \hat{\omega}^{I\beta}_{\mathrm{SB}}: \mathfrak{B}(\mathfrak{H}_\mathrm{SB})\rightarrow \mathbb{C}
\end{equation} \textit{be constructed as in section} \ref{sectionX} \textit{and let $\mathfrak{L}_{ t}^\beta: \mathfrak{A}_\Lambda\rightarrow\mathfrak{B}(\mathfrak{H}_\mathrm{M})$, $\mathfrak{A}_\Lambda\ni A\mapsto\mathfrak{L}_{ t}^{ \beta}(A)\in\mathfrak{B}(\mathfrak{H}_\mathrm{M})$ be given by} 
\begin{equation}\label{eq:L}
\mathfrak{L}_{ t}^{ \beta}(A):\mathbb{R}^{3K}\times\mathbb{R}^{3E}\rightarrow \mathbb{C}\quad with\quad \mathbb{R}^{3K}\times\mathbb{R}^{3E}\ni (X,X^\mathrm{e})\mapsto \mathfrak{L}_{ t}^{ \beta}(A)(X,X^\mathrm{e})\,\equiv\,\hat{\omega}^{I\beta}_{\mathrm{SB}}(\pi_\mathrm{SB}^\beta\circ\tau_{Pt}^{I\mathrm{SB}}(A))(X,X^\mathrm{e}).
\end{equation} \textit{We calculate $\hat{\omega}^{I\beta}_{\mathrm{SB}}$ according to eq. (\ref{eq:ESBS}) and $\tau_{Pt}^{I\mathrm{SB}}(A)$ according to eq.(\ref{eq:SBDynamics}), by using $H^\mathrm{SB}_\mathrm{Int}+P_t$ instead of $H^\mathrm{SB}_\mathrm{Int}$. For a molecular system described by $\omega^\beta_\mathrm{M}$ according to eq. (\ref{eq:MolecularSystem}) the NMR signal $\langle M^+ \rangle_\beta(t)$ is defined by} 
\begin{equation}\label{eq:NMRsignal}
 \langle M^+ \rangle_\beta(t)\,\dot{=}\,\sum_{j=1}^K\omega^\beta_\mathrm{M}\big(\mathfrak{L}_{ t}^{ \beta}( I^+_j)\big).
 \end{equation} \textit{A reconstruction of the probability density $\mid \Psi^\beta(X) \mid^2$ for the spatial distribution of the nuclei is achieved by identifying an $\omega^\beta_\mathrm{M}$ which provides a sufficient agreement between the calculated and the experimental NMR spectrum.}
\medskip \\
The identification of $\omega^\beta_\mathrm{M}$ may be based on an initial guess (e.g. by inserting the PES into the classical Gibbs state) with subsequent manual adaptions. 
Note that $\mathfrak{L}_{ 0}^\beta$ gives thermal equilibrium at the beginning of the experiment and $\mathfrak{L}_{t>0}^\beta$ describes the time-evolution during the experiment. 
 The notation $\mathfrak{L}_{ t}^{ \beta}(A)(X,X^\mathrm{e})$ is unconventional but easier to read in later applications. A conventional notation is given by $\mathfrak{L}_{\beta t}^{A}(X,X^\mathrm{e})\equiv\mathfrak{L}_{ t}^{ \beta}(A)(X,X^\mathrm{e})$ but this is more difficult to read when dealing with $\omega^\beta_\mathrm{M}\big(\mathfrak{L}_{\beta t}^{ I^+_j}\big)$. As usual the (1-dimensional) NMR spectrum, $S_\beta(\nu)$, is calculated as the Fourier transform
\begin{equation}\label{eq:NMRspectrum}
S_\beta(\nu)=\int^\infty_0dt\, \langle M^+\rangle_\beta(t)e^{-i\nu t}.
\end{equation}
The structural validity of the main result will now be checked in the next two sections.

\section{Breakup of the effective spin model}
The breakup of the effective spin model is shown for the time-independent expectation value in thermal equilibrium as well as for the out of equilibrium spin dynamics. In thermal equilibrium the expectation value of the z-component of a nuclear spin of a molecule is reduced, if compared to the case where the spin is isolated. This is due to the action of the external magnetic field on the magnetic moments of the surrounding electrons, which then reduce the external magnetic field at the positions of the nuclei. For diagonal $\sigma_j$ the effective model from eq. (\ref{eq:Eff}) provides
\begin{equation}\label{eq:ExpEff}
\langle I^\mathrm{z}_j\rangle_\mathrm{eff} = \mathrm{Tr}(\rho^\beta_\mathrm{eff}  I^\mathrm{z}_j)\approx \frac{\hbar^2}{4}\beta\gamma_jB^\mathrm{z}_{\mathrm{ext}}
(1-\sigma^{\mathrm{zz}}_j) \quad \mathrm{where} \quad \rho^\beta_\mathrm{eff} =\frac{\exp(-\beta H_\mathrm{eff})}{\mathrm{Tr}(\exp(-\beta H_\mathrm{eff}))}.
\end{equation} 
Higher order terms can be neglected in the high temperature approximation. The effective magnetic shielding (chemical shift) constant is always small and positive, i. e., $1\gg\sigma^{\mathrm{zz}}_j>0$. Hence, the expectation value of an isolated nuclear spin is reduced in the molecular system by $\sigma^{\mathrm{zz}}_j$.

In this document the hydrogen atom is used as basic example for mQED calculations. One finds similar results for a Helium atom. Remember that $I^\mathrm{z}$ denotes the z-component of the spin operator of the proton while $S^\mathrm{z}$ denotes the operator from the electron. While the first order of eq. (\ref{eq:ESBS}) is zero for $A=I^\mathrm{z}$ one derives in the second order that
\begin{align}\label{eq:Limit}
\omega^\beta_\mathrm{M}\big(\mathfrak{L}_0^\beta(I^\mathrm{z})\big)=&
 \,\omega^\beta_\mathrm{S}( I^\mathrm{z})-  \omega^\beta_\mathrm{S}(S^\mathrm{z})\, r_{\varphi}^\beta+... \\
\label{eq:Limit2} \approx& \frac{\hbar^2}{4}\beta\gamma B^\mathrm{z}_{\mathrm{ext}} (1-a_{\varphi}^\beta).
 \end{align} The dots (...) denote higher order terms from the perturbation series. $r_{\varphi}^\beta$ and $a_{\varphi}^\beta$ differ by a constant and the high temperature approximation is made for $\omega^\beta_\mathrm{S}( I^\mathrm{z})$ and $\omega^\beta_\mathrm{S}( S^\mathrm{z})$. It can be seen that $a_{\varphi}^\beta$, derived \textit{non}-effectively from mQED, replaces the effective parameter $\sigma^{\mathrm{zz}}_j$ which is commonly derived according to eq. (\ref{eq:Ramsey}). It can be checked that $a_{\varphi}^\beta$ is dimensionless and therefore $a_{\varphi}^\beta$ can be given in "parts per milion" (ppm) in analogy to $\sigma^{\mathrm{zz}}$. One finds
 \begin{equation}\label{eq:a}
 a_{\varphi}^\beta=\frac{g^2_s\mu_\mathrm{B}^2}{4}\int^\beta_0ds_1\int^{s_1}_0ds_2 \iiint_{\mathrm{R}^3} d^{3}x\iiint_{\mathrm{R}^3} d^{3}x^\mathrm{e}
  \iiint_{\mathrm{R}^3} d^{3}k \mid \Psi^\beta(\vec{x}) \mid^2 \mid \psi_{100}(\vec{x},\vec{x^\mathrm{e}})\mid^2 \mathfrak{m}^{\mathrm{zz}}_{\varphi\beta}(\vec{x},is_2,\vec{x^\mathrm{e}},is_1,\vec{k}).
\end{equation} In case of a hydrogen atom $a_{\varphi}^\beta$ is indeed independent from a particular choice of the nuclear wave function $\Psi^\beta$. This reflects the fact that the magnetic shielding is independent from the position of the atom in the homogeneous external field. The distribution of the electron is chosen to be the 1-s orbital of the hydrogen atom, i. e.
\begin{equation}
\psi_{100}(\vec{x})=\frac{1}{\sqrt{\pi a_\mathrm{B}^3}} \,\exp{\bigg(\frac{-\mid\vec{x}-\vec{x^\mathrm{e}} \mid}{a_\mathrm{B}}\bigg)},
\end{equation} where $a_\mathrm{B}$ is the Bohr radius.
\medskip \\
\textbf{Observation 1}
\medskip \\
A comparison provides a further advantage for the non-effective model. For $B^\mathrm{z}_{\mathrm{ext}}\rightarrow\infty$ we have $\sigma^{\mathrm{zz}}B^\mathrm{z}_{\mathrm{ext}}\rightarrow\infty$. Hence, the effective model predicts that
the magnetic field which originates from the electron and reduces the magnetic field at the position of the proton tends to infinity. This is certainly wrong because there is a maximum magnetic field strength which can be produced by the electron and the maximum is achieved when the spin of the electrons is completely in the $\mid+1/2\rangle$ or $\mid-1/2\rangle$ state. In contrast the mQED model contains this effect and the limit is given by 
$\mathrm{Tr}(\rho^\beta_2 S^\mathrm{z})\leq\hbar/2$ in eq. (\ref{eq:Limit}). In this case the high temperature approximation made in $a_{\varphi}^\beta$ is unsuitable and eq. (\ref{eq:Limit}) provides more accurate predictions than eq. (\ref{eq:Limit2}). Thus, for low temperatures and high external magnetic fields the mQED model is much more realistic. The deviation from the non-linear regime for $\sigma^{\mathrm{zz}}B^\mathrm{z}_{\mathrm{ext}}\rightarrow\infty$ may be measured experimentally and validates the more realistic description of the mQED model. This is of potential relevance for NMR at low temperatures, e. g. Dynamic Nuclear Polarization (DNP).
\medskip \\
\textbf{Observation 2}
\medskip \\
We have $a_{\varphi}^\beta>0$ for all $\varphi\in L^2(\mathbb{R}^3)$ which follows from the fact that $\varphi$ enters $a_{\varphi}^\beta$ with 
$\mid\varphi(\vec{k})\mid^2$ and 
\begin{equation}\label{eq:Shielding}
a_{\varphi}^\beta = \frac{g^2_s\mu_\mathrm{B}^2\mu_0}{6\pi^{2} \hbar c}\int_0^\infty dk \mid\varphi(k)\mid^2\underbrace{\bigg(\frac{e^{-2\beta\hbar ck}}{2}-e^{-\beta\hbar ck} +0.5\bigg)}_{>0\forall k\in\mathbb{R}^+}\bigg(1+2\rho^\beta(k) \bigg)\frac{k}{(1+k^2a_\mathrm{B}^2/4)^2}.
\end{equation} In eq. (\ref{eq:Shielding}) spherical coordinates were introduced for $\vec{k}$ and all integrals except the one for $k$ were evaluated. Furthermore, we have assumed that $\varphi$ depends only on $\mid\vec{k}\mid$ which is a natural and common choice. This is a nice result, because the non-effective magnetic shielding $a_{\varphi}^\beta$ needs to be positive is any case and $\varphi$ is a free parameter in mQED. 
\medskip \\
\textbf{Numerical investigation of $a_{\varphi}^\beta$.}
\medskip \\
It is important to know that for any given molecular structure the coupling function $\varphi$ from eq. (\ref{eq:MagneticField}) is the only free parameters and - of course - $\varphi$ is independent of $\Psi$ and $\psi$. Thus, a particular choice for $\varphi$ which accurately reproduces a well-understood experiment can be used to predict or analyze NMR data of proposed or unknown molecular structures. We have $g^2_s\mu_\mathrm{B}^2\mu_0/6\pi^{2} \hbar c\approx 7.271326950237399 \cdot 10^{-8} \mathrm{\r A}^2$, where $\mathrm{\r A}$ is the unit $\mathrm{\r A}$ngstr\"{o}m. For the numerical calculations we choose
\begin{equation}
\varphi(k)=\bigg\{\begin{array}{l} g\quad \mathrm{for}\quad \delta_{\mathrm{IR}}\leq k \leq  \delta_{\mathrm{UV}} \\ 
0 \quad \mathrm{for}\quad k <\delta_{\mathrm{IR}} \,\vee\, k> \delta_{\mathrm{UV}}  
\end{array}
\quad\quad g,\delta_\mathrm{UV}\in\mathbb{R}^+ \quad\mathrm{and}\quad \delta_{\mathrm{UV}} >\delta_\mathrm{IR}\in\mathbb{R}^+_0.
\end{equation} $\delta_{\mathrm{IR}}$ and $\delta_{\mathrm{UV}}$ are the infrared and ultraviolett cutoff respectively and normalization of $\varphi$ implies $g=1/(\delta_{\mathrm{UV}}-\delta_{\mathrm{IR}})$. For the numerical calculations in this document the infrared cutoff $\delta_{\mathrm{IR}}$ can be chosen arbitrarily small (and also zero) such that it has a negligible influence on $a_{\varphi}^\beta$. 

\begin{figure}[h] 
\centering
\includegraphics*[scale=0.95]{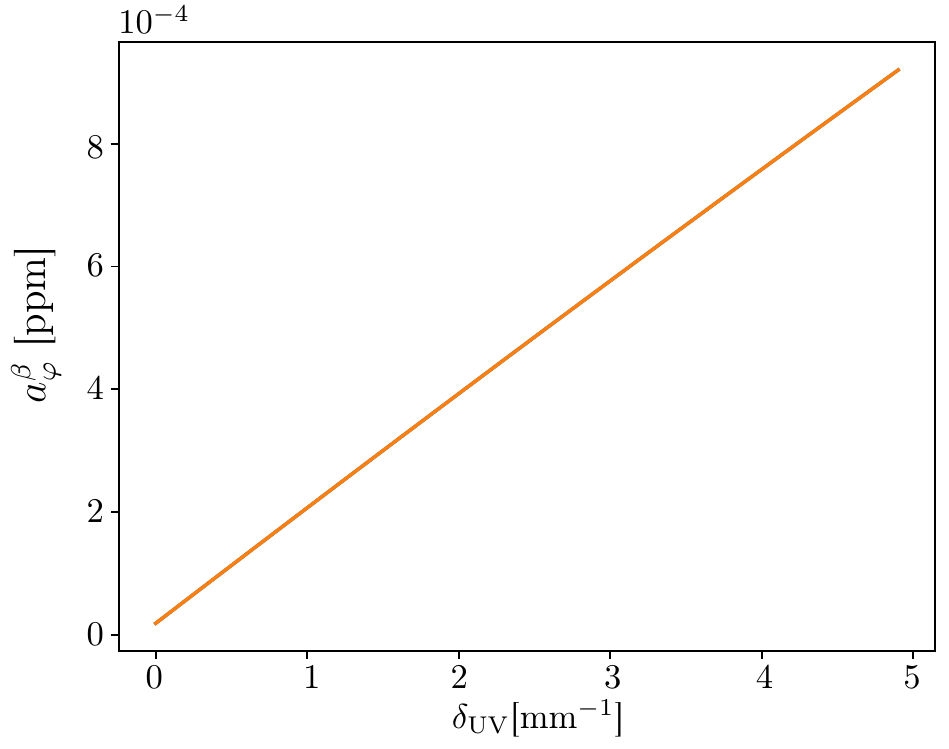}
\caption{\label{fig:Shielding} The magnetic shielding $a_{\varphi}^\beta$ is shown as a function of $\delta_{\mathrm{UV}}$. It can be seen that there is a linear dependence which is in agreement with the dynamical calculations shown below.}
\end{figure}

Figure (\ref{fig:Shielding}) shows the dependence of $a_{\varphi}^\beta$ on $\delta_{\mathrm{UV}}$ for a hydrogen atom. The unit of $\delta_{\mathrm{UV}}$ is millimeter$^{-1}$ (1 mm$^{-1}$ $\approx$ 0.00124 eV). The temperature is chosen to be T=293 K (room temperature) and the infrared cutoff is chosen to be zero, $\delta_{\mathrm{IR}}=0$. There is a linear increase of $a_{\varphi}^\beta$ for increasing $\delta_{\mathrm{UV}}$ which is in agreement with the dynamical calculations shown below. The ppm (parts per million) scale is chosen such that the Lamor frequency $v_0$ is located at zero. Higher order contributions should increase the magnetic shielding.
\medskip \\
\textbf{The dynamic case.}
Analytically the breakup of the effective model can, for example, be seen by the occurrence of direct spin-spin interactions (dipole-dipole interactions) in the second order of eq. (\ref{eq:SBDynamics}). One finds,
\begin{equation}
[\tau^{\mathrm{SB}}_{t_2}(H^\mathrm{SB}_\mathrm{Int}),[\tau^{\mathrm{SB}}_{t_1}(H^\mathrm{SB}_\mathrm{Int}),\tau^{\mathrm{SB}}_{t}(I^+)]]
=\sum_{i< j}J_{ji}^{\mathrm{zz}}(t,t_1,t_2)\,I^\mathrm{z}_j\otimes I^\mathrm{z}_i+...
\end{equation} with  
\begin{equation}
J_{ji}^{\mathrm{zz}}(t,t_1,t_2)=-\gamma_j\gamma_i u_j(t)\bar{u}_j(t_1) \big([B^\mathrm{z}_{0\varphi}(\vec{x_i},t_2),B^\mathrm{x}_{0\varphi}(\vec{x_j},t_1)]+i[B^\mathrm{z}_{0\varphi}(\vec{x_i},t_2),B^\mathrm{y}_{0\varphi}(\vec{x_j},t_1)] \big).
\end{equation} and similar terms for the x- and y-components. Following the calculations from \cite{Dipole} the direct coupling $\vec{I}_iD_{ij}\vec{I}_j$ from the effective model eq. (\ref{eq:Eff}) is obtained with quantum radiative corrections. Indirect spin-spin couplings occur in the fourth order of eq. (\ref{eq:SBDynamics}) in a similar fashion. For the magnetic shielding in the dynamic case the numerical investigation of the breakup of the effective spin model is detailed shown in the next section.

\section{NMR spectra from molecular Quantum Electrodynamics at finite temperatures.}
The real-time nuclear spin dynamics as well as the spectra according to eq. (\ref{eq:NMRspectrum}) are calculated in the second order of eq. (\ref{eq:SBDynamics}) and the second order of eq. (\ref{eq:ESBS}) according to eq. (\ref{eq:NMRsignal}) and eq. (\ref{eq:NMRspectrum}). After long-lasting calculations NMR-spectra are obtained from terms of the form
\begin{align}\label{eq:S}
S_\beta(\nu)&=\vartheta_1\int^\infty_0dt e^{-i\nu t} \int^\beta_0ds_1\int^{s_1}_0ds_2 \int^t_0dt_1\int^{t_1}_0dt_2 \iiint_{\mathrm{R}^3} d^{3}x\iiint_{\mathrm{R}^3} d^{3}x^\mathrm{e}
  \iiint_{\mathrm{R}^3} d^{3}k   \iiint_{\mathrm{R}^3} d^{3}k' \times \\
  \nonumber&\times\mid \Psi^\beta(\vec{x}) \mid^2\mid \psi_{100}(\vec{x},\vec{x^\mathrm{e}})\mid^2  \times \\
  \nonumber&\nonumber\times\big(  \mathfrak{m}^{\mathrm{xz}}_{\varphi\beta}(\vec{x},is_2,\vec{x^\mathrm{e}},is_1,\vec{k}')\, \omega^\beta_\mathrm{S}(I^\mathrm{y}\tau_{is_2}(I^\mathrm{x}))+ \mathfrak{m}^{\mathrm{yz}}_{\varphi\beta}(\vec{x},is_2,\vec{x^\mathrm{e}},is_1,\vec{k}')\, \omega^\beta_\mathrm{S}(I^\mathrm{y}\tau_{is_2}(I^\mathrm{y})) \big)\times \\
  \nonumber&\times \sum_{\lambda=1,2}\varphi^\mathrm{z}_\lambda(\vec{k})\big(\varphi^\mathrm{x}_\lambda(\vec{k})\, +i\varphi^\mathrm{y}_\lambda(\vec{k})\,\big)i\Delta_{\vec{k}}((\vec{x^\mathrm{e}},t_2)-(\vec{x},t_1))
 u(t)\bar{u}(t_1) +...
\end{align} where $u(t)=e^{-itv_0}$ and $\vartheta_1\in\mathbb{C}$. In case of the hydrogen atom one finds again that $S_\beta(\nu)$ is independent of $\Psi^\beta$ which means that the chemical shift does not depend on the position of the hydrogen atom in the homogenous external field. For molecules with two or more nuclei $S_\beta(\nu)$ depends on $\Psi^\beta$.
 
\begin{figure}[h]
\centering
\includegraphics*[scale=0.95]{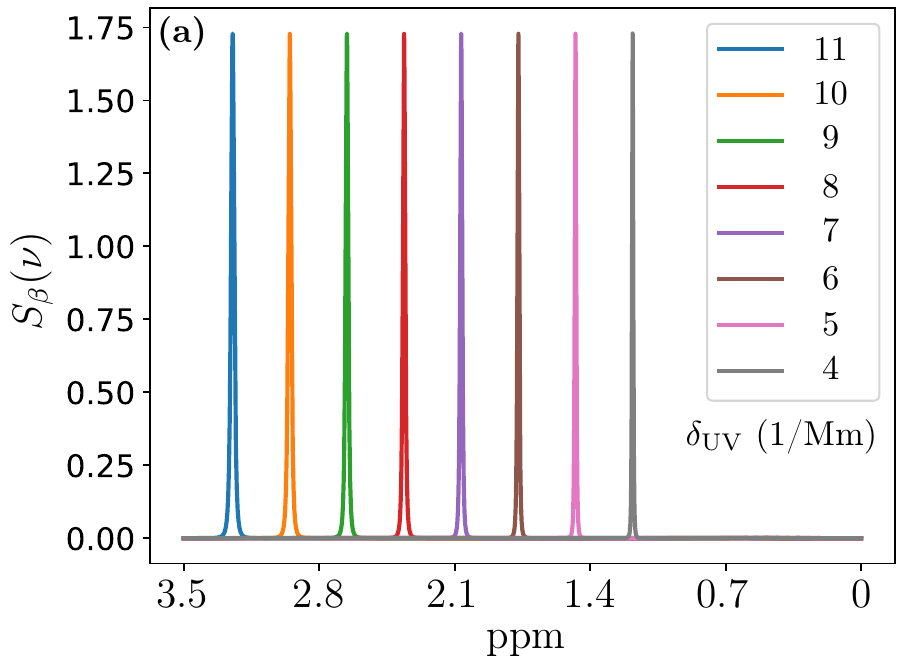}
\includegraphics*[scale=0.95]{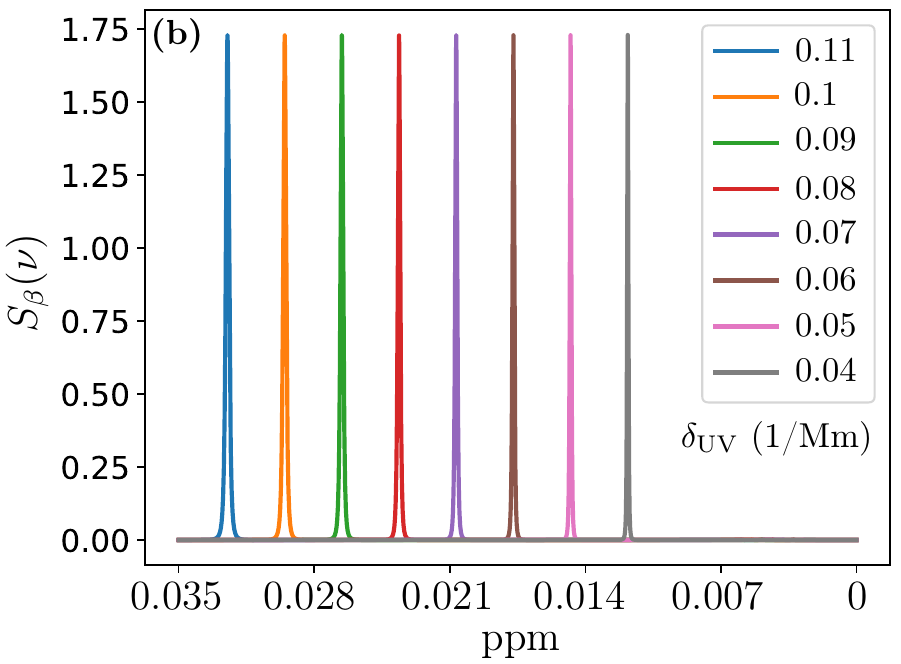}
\caption{\label{fig:SpectraUV} NMR-spectra $S_\beta(v)$ of a hydrogen atom calculated from eq. (\ref{eq:NMRsignal}) with eq. (\ref{eq:NMRspectrum}) for different values of $\delta_\mathrm{UV}$. All peaks have a Lorentz shape which is in agreement with experimental data. The chemical shift and $\Delta\nu$ increase linear with $\delta_\mathrm{UV}$.}  
\end{figure}

 Figure \ref{fig:SpectraUV} shows the NMR spectra $S_\beta(v)$ from eq. (\ref{eq:NMRsignal}) with eq. (\ref{eq:NMRspectrum}) for a hydrogen atom. The ppm scale is chosen such that the Lamor frequency $v_0$ is located at zero ppm. The spectra is calculated for the values $\delta_{\mathrm{UV}}= 4, 5, 6, 7, 8, 9, 10, 11$ (a) and $\delta_{\mathrm{UV}}= 0.04, 0.05, 0.06, 0.07, 0.08, 0.09, 0.01, 0.011$ with unit megameter$^{-1}$ (Mm$^{-1}$). The temperature is chosen to be $T=293$ K (room temperature), $ B^{\mathrm{z}}_\mathrm{ext}=20$ T (Tesla) and the infrared cutoff is chosen to be zero, $\delta_{\mathrm{IR}}=0$. In every case it can be seen that a Lorentz distribution is obtained as observed in NMR experiments. Small variations of $\delta_{\mathrm{IR}}$ only had a negligible impact on the magnetic shielding.  As in the case for $a^\beta_{\varphi}$ the strength of the magnetic shielding increases linear with $\delta_\mathrm{UV}$. Furthermore, the "Full Width at Half Maximum" (FWHM) $\Delta\nu$ increases linearly with increasing $\delta_\mathrm{UV}$. Remember that $\Delta\nu$ is directly related with the life-time of an excited nucleus which will be checked later. Comparing the left figure (a) and the right figure (b) one finds that if $\delta_\mathrm{UV}$ is reduced by a factor of 100 then $\Delta\nu$ as well as strength of the magnetic shielding (distance of the peak to 0 ppm) is also reduced by a factor of 100. The maximum value (height) of each peak is nearly the same. This a result of the normalization of $\varphi$ and it makes sense because these small changes of the magnetic field strength should not have a significant impact on the amplitude of the NMR signal. This is also in agreement with experimental data.

\begin{figure}[h]
\centering
\includegraphics*[scale=0.8]{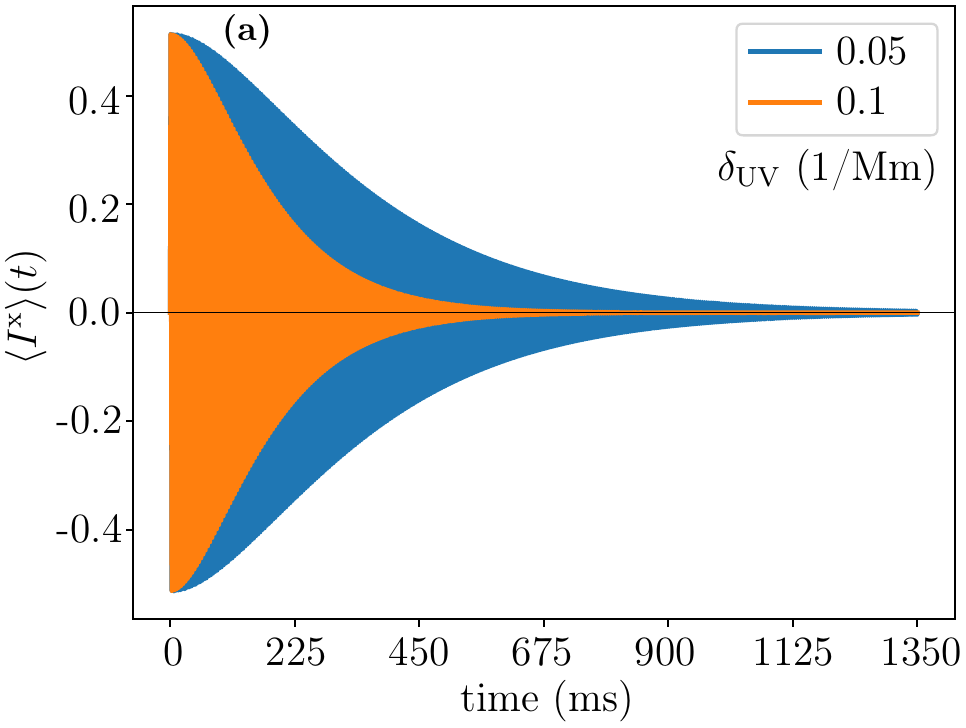}
\includegraphics*[scale=0.8]{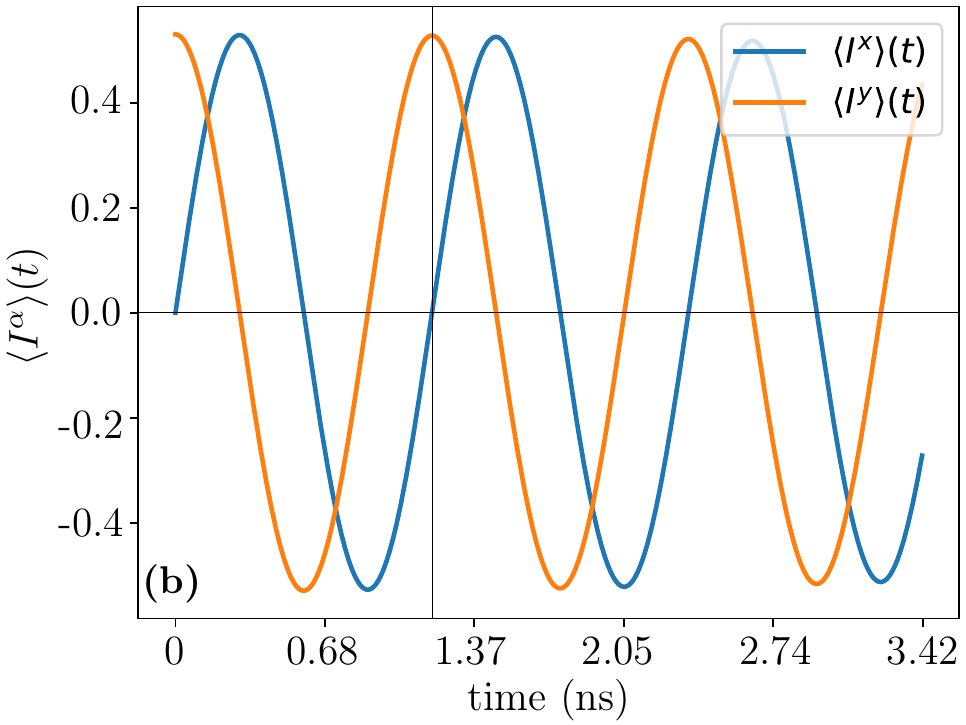}
\caption{\label{fig:Dynamics} (a) Long-time spin dynamics from the real part of the NMR-signal, $\Re (\langle M^+ \rangle_\beta(t))=\langle I^\mathrm{x}\rangle (t)$, for $\delta_\mathrm{UV}=0.1$ Mm (orange line) and $\delta_\mathrm{UV}=0.05$ Mm (blue line). Doubling the value $\delta_\mathrm{UV}$ halves the life-time of the NMR-signal. (b) Short time spin dynamics $\langle I^\mathrm{y}\rangle (t)$ (orange line) and $\langle I^\mathrm{x}\rangle (t)$ (blue line) for the same parameters as used for the orange line from fig. (\ref{fig:Dynamics} a). There is an exact $90^{\circ}$ phase shift between  $\langle I^\mathrm{y}\rangle (t)$ and $\langle I^\mathrm{x}\rangle (t)$ as it should be. The very smooth exponential decay is due to the fact that the NMR signal contains only one peak. Several peaks provide oscillations as in fig. (\ref{fig:ExpData}).}
\end{figure}

Figure \ref{fig:Dynamics} (a) shows the long-time dynamics of $\langle I^\mathrm{x}\rangle (t)$ (real part of NMR-signal) for $\delta_\mathrm{UV}=0.1$ Mm (orange line) and $\delta_\mathrm{UV}=0.05$ Mm (blue line). All other parameters are the same as for the calculations for fig. \ref{fig:SpectraUV}. In both cases there is an exponentially fast return to equilibrium as observed in NMR experiments. The starting point at $t=0$ is chosen to be directly after the $90^{\circ}$-pulse has finished. The amplitudes are normalized to the value 0.5 at $t=0$ corresponding to the excitation of a single nucleus. The thermalization which is associated with the orange line happens twice as fast as the thermalization which is associated with the blue line. Hence, doubling the value $\delta_\mathrm{UV}$ halves the life-time (T2 in NMR language) of the excited spin. The nuclear spin can release energy in a frequency range with double length.

Figure \ref{fig:Dynamics} (b) shows the short-time dynamics $\langle I^\mathrm{y}\rangle (t)=\Im (\langle M^+ \rangle_\beta(t))$ and $\langle I^\mathrm{x}\rangle (t)=\Re(\langle M^+ \rangle_\beta(t))$ for the same parameters which were used for the orange line from fig. \ref{fig:Dynamics} (a). The cross shows that there is an exact $90^{\circ}$ phase shift between $\langle I^\mathrm{y}\rangle (t)$ and $\langle I^\mathrm{x}\rangle (t)$ as it should be. The frequency is slightly reduced compared to the Lamor frequency which can also be seen from fig. \ref{fig:SpectraUV} (b).

\section{Outlook}\label{sec:Outlook}
\begin{figure}[h]     
\centering
\includegraphics*[scale=1]{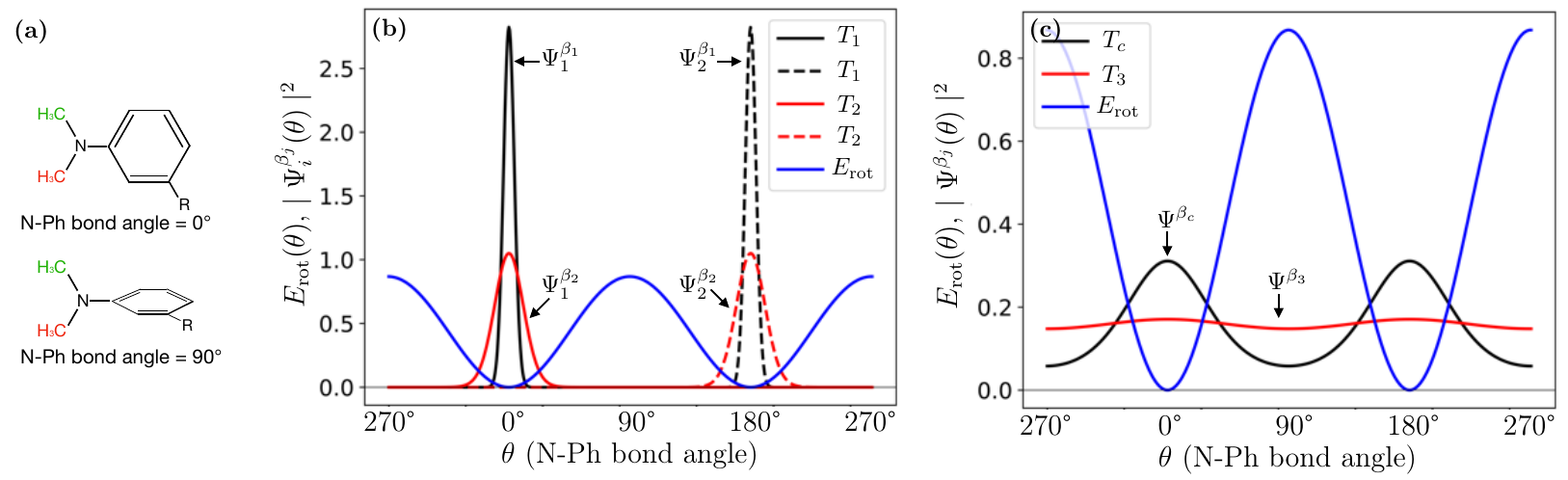}
\caption{\label{fig:2} In mQED \textit{all} bond angles of a molecule (a) are considered with a continuous probability distribution $\mid \Psi^\beta(X)\mid^2$, which depends on the temperature ((b) and (c)). For each bond angle $\theta$ there is an energy $E_\mathrm{rot}(\theta)$ which often has a similar form like the blue line in (b) and (c). The approximation $\rho^\beta(E_\mathrm{rot}(\theta))=\mid \Psi^\beta(\theta)\mid^2$ provides the following: For relatively low temperatures $T_1$ compared with $E_\mathrm{rot}$, the probabilities are such that mainly bond angles with lower energy are populated (black line $\Psi^{\beta_1}_1$ or dashed black line $\Psi^{\beta_1}_2$ in (b)). If the temperature increases ($T_2$), then neighboring bond angles become more and more occupied (red line $\Psi^{\beta_2}_1$ or dashed red line $\Psi^{\beta_2}_2$ in (b)). If the temperature increases increases even more ($T_c)$, also bond angles with highest energy become significantly occupied (black line $\Psi^{\beta_c}$ in (c)). Finally, if the temperature $T_3$ is high, then all bond angles become nearly equally populated (red line $\Psi^{\beta_3}$ in (c)). This is due to the maximization of the entropy.} 
\end{figure}

In this section it is outlined how the new approach can serve for a more detailed molecular structure determination compared to conventional NMR theory. To obtain an approximated amplitude square $\mid\Psi^\beta(X)\mid^2$ of the nuclear KMS state, the potential energy surface (PES), $E_\mathrm{PES}(X)$, or rotational energies, $E_\mathrm{rot}(\theta)$, from Quantum Chemistry may be inserted in to the classical Gibbs state $\rho^\beta$ at inverse temperature $\beta$ (fig. \ref{fig:2}). 

\begin{figure}[h]     
\centering
\includegraphics*[scale=0.86]{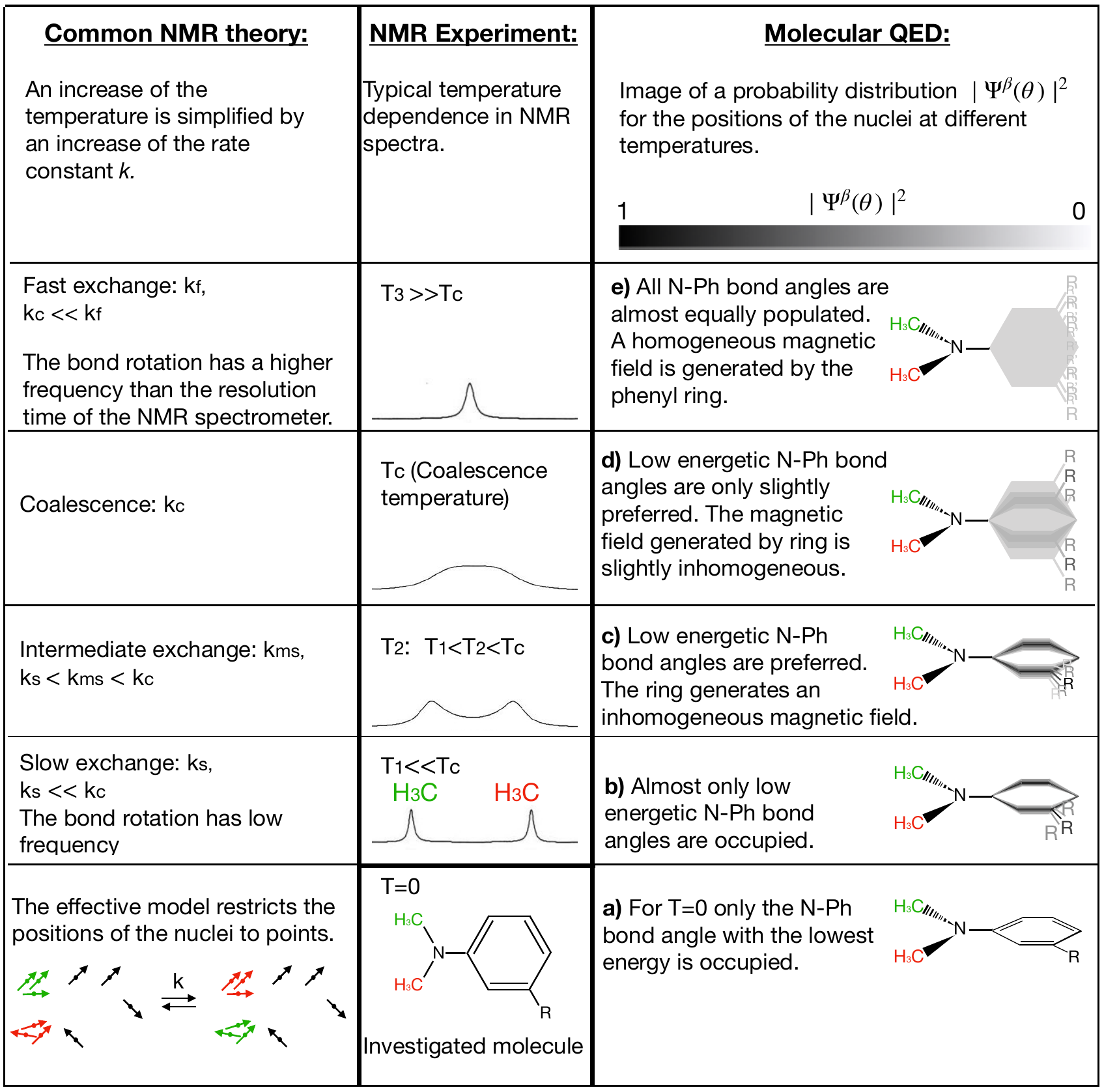}
\caption{\label{fig:3} The influence of the temperature on a molecular structure can often be observed in experimental NMR data (middle column).
In conventional NMR theory, the positions of the nuclei are restricted to fixed points while the influence of the temperature on the molecular structure is phenomenologically simplified and included in a rate constant $k$ (left column).  Hence, the molecular structure is only roughly approximated in conventional NMR theory. In contrast, mQED allows an inclusion of the more detailed probability density $\mid \Psi^\beta(\theta)\mid^2$ for the spatial distribution of the nuclei (right column). This heuristic illustration serves as motivation for the presented method. However, the structural validity of this initially heuristic explanation is shown in section VI and VII. }
\end{figure}

From Quantum Statistical Mechanics we know that for different temperatures $T_1<T_2<T_c<T_3$ there are different probabilities $\mid \Psi^\beta(\theta)\mid^2$ for the molecule to have a certain bond angle $\theta$ (fig. \ref{fig:2}). Such effects can be observed in NMR, because the green and the red methyl groups (a) can have different electronic environments which depends on the temperature. The probability $\mid \Psi^\beta(\theta)\mid^2$ is time-independent in chemical and thermodynamic equilibrium such that the molecule is in a superposition of several bond angels. This is in contrast to conventional NMR theory, where the N-Ph bond rotates with a certain frequency. The probability distribution $\rho^\beta(E_\mathrm{rot}(\theta))$ from fig. (\ref{fig:2}) is only a rough approximation for the more realistic, quantum mechanical probability distribution $\mid \Psi^\beta(\theta)\mid^2$. Hence, in a second second step, the distribution $\rho^\beta$ can be slightly changed until the calculated NMR spectra agree with experimental NMR data. Hence, the probability distribution $\mid \Psi^\beta(X)\mid^2$ can be reconstructed from NMR data by using mQED at finite temperatures. A significant advantage of mQED is that the impact of the temperature on the molecular structure is taken into account much more realistically compared to conventional NMR theory. The result is that more realistic and more detailed molecular structures may be decoded from experimental NMR data. As a motivation for the presented method a heuristic illustration for a more detailed structure determination is outlined in fig. \ref{fig:3}. NMR spectra of such a molecule are not calculated in this document, but structural validity is shown.
 
\begin{figure}[h]     
\centering
\includegraphics*[scale=0.88]{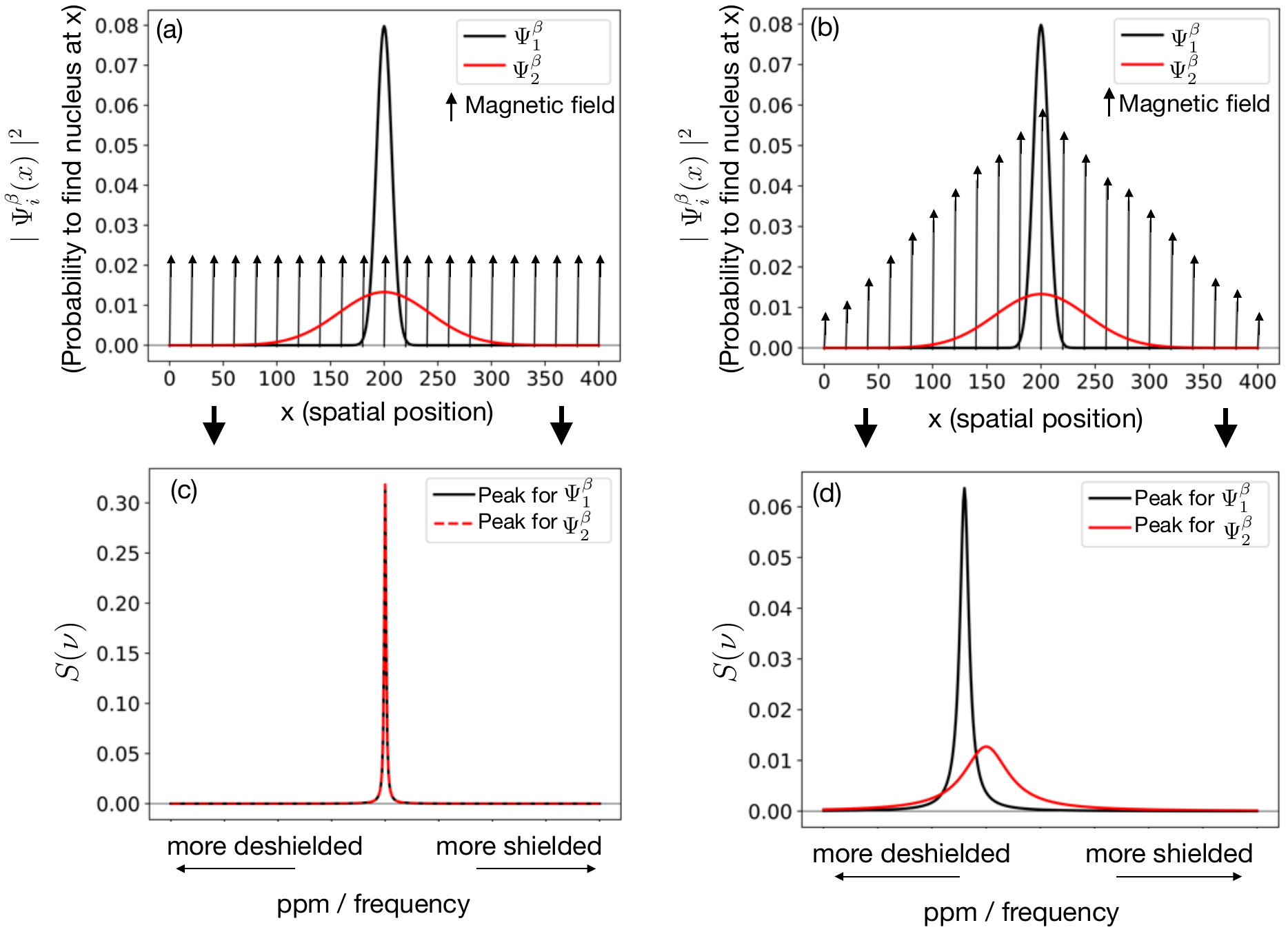}
\caption{\label{fig:4} Two different spatial probability distributions $\mid \Psi_1^\beta(X)\mid^2$ and $\mid \Psi_2^\beta(X)\mid^2$  are placed in a homogenous magnetic field (a) and in an inhomogeneous magnetic field (b). In mQED, the NMR spectrum $S(\nu)$ is independent from the spatial distribution of the nuclei if the magnetic field is homogenous (c). If the magnetic field depends on position in space (because of the distribution of electrons), then different spatial distributions of the nuclei generate (in general) different NMR peaks (d). This schematic (and not precise) illustration will be mathematically verified in section VI and VII.}
\end{figure}

In the zero temperature limit only the bond angle with the lowest energy is occupied (lowest row, (a) in fig. \ref{fig:3}). Such ground state structures are obtained from common quantum chemical calculations (like DFT). At low temperatures (second row from below, (b) in fig. \ref{fig:3}), where mainly low energetic bond angles are occupied (fig. \ref{fig:2}), the magnetic field generated by the ring is different for the green and the red marked methyl group. Hence, both methyl groups have clearly distinct NMR signals. In contrast, conventional NMR theory assumes that the ring is slowly rotating with a fixed frequency. Hence, mQED and common NMR theory provide two different structures for the same situation. In mQED, bond angles with higher energies become more and more occupied with increasing temperature (third row from below, (c) in fig. \ref{fig:3}). Hence, each of the methyl groups comes closer to the opposite side of the ring. As a result both peaks on the spectrum come closer to each other. In conventional NMR theory, the rotation frequency is just slightly enhanced. However, in mQED there are still some bond angles which are nearly unoccupied ((b) in fig. \ref{fig:2} and third row from below, (c) in fig. \ref{fig:3}). In NMR there is a specific temperature called "coalescence temperature" $T_c$, where the two peaks start to merge. At this temperature, also bond angles with higher energy are occupied but lower energetic bond angles are still preferred in mQED (black line in (c) of fig. \ref{fig:2} and (d) in fig. \ref{fig:3}). This is not contained in conventional NMR theory, where the rotation frequency $k$ is simply increased. At high temperature the two peaks merge completely and provide one sharp peak in the experimental NMR spectrum. In the interpretation of conventional NMR theory, the rotation frequency is much higher than the temporal resolution of the NMR spectrometer such that only one averaged signal is observed. The interpretation in mQED is that nearly all bonding angles are equally occupied in order to maximize the entropy. In this case, the resulting magnetic field is for both methyl groups the same, because it is spatially averaged. Hence, both methyl groups have the same chemical shift. The structural validity of this initially heuristic explanation is underpinned by an illustration (fig. \ref{fig:4}) which is mathematically verified in section VI and VII. 

In a region where the magnetic field is homogeneous the NMR peaks are independent from the spatial distribution of the nuclei ((a) and (c) in fig. \ref{fig:4}). Note that for high temperatures the distribution associated with $T_3$ in fig. \ref{fig:2} generates a homogenous magnetic field in the region of both methyl groups ((e) in fig. \ref{fig:3}). Hence, both peaks are equal. However, if the magnetic field is inhomogeneous in the region of the methyl groups, because the spatial probability distribution depends on the bond angle (fig. \ref{fig:2}, \ref{fig:3} and \ref{fig:4}), the methyl groups do not generate the same peak in the NMR spectrum $S(\nu)$. For relatively low temperatures the methyl groups are strongly localized in an inhomogeneous field. The NMR peaks will broaden with increasing delocalization of the methyl groups in an inhomogeneous magnetic field (fig. \ref{fig:4}) and hence have clearly distinct peaks. At the coalescence temperature $T_c$ there are delocalized methyl groups in an inhomogeneous magnetic field. With increasing temperature the magnetic field generated from the ring becomes more and more homogenous until both peaks merge to a single sharp peak at $T_4$.

\section{Conclusion} 
In this document it was shown how NMR spectra are mathematically connected with the quantum statistical and temperature-dependent probability density $\mid\Psi^\beta(X) \mid^2$ for the spatial distribution of the nuclei. Thus, in the presented method the nuclei can be continuously distributed in position space $\mathbb{R}^3$. 
This improves the concept of current NMR theory in which the nuclei are restricted to fixed points in a lattice $L$, while the omnipresent delocalization of nuclei is either phenomenologically simplified in the form of rate constants or completely neglected. Furthermore, the temperature can significantly influence NMR spectra and the thermal energy has an important impact on the molecular structure. 
However, the effective spin model, which is the basis for conventional NMR theory, is almost independent of the temperature. This weak point is removed in the new method as well.
Hence, the presented method provides a foundation for a more realistic and more detailed determination of molecular structures.
 
The main result (page 11) provides the structural application of molecular Quantum Electrodynamics and Quantum Statistical Mechanics in the algebraic reformulation to Nuclear Magnetic Resonance. Analytical and numerical calculations as well as comparisons with experimental NMR data showed the validity of this approach. Furthermore, wrong predictions of the effective spin model are corrected by the new approach (observation 1) and several striking advantages against established NMR theory were discussed. The presented method makes use of the physical approximation that the energy of an NMR pulse is too weak to change the molecular geometry which is also used in the effective spin model and obviously realistic for NMR. The important process of return to equilibrium is included in a natural and microscopic way instead phenomenologically as in eq. (\ref{eq:effRelax}). This provides a basis for a more detailed research towards optimized polarization transport and stable spin structures which are of basic interest in hyperpolarized MRI. Chemical shifts (magnetic shieldings) as well as spin-spin couplings occur naturally and must not be described effectively. Hence, quantum radiative corrections are naturally included in the calculated NMR spectrum.  

The fundamental problem of performing numerical calculations with the infinite-dimensional radiation field at finite temperatures was solved by
using a purified version of the Araki-Woods representation which served as a key element. The perturbation series eq. (\ref{eq:SBDynamics}) in combination with eq. (\ref{eq:ESBS}) generates combinations of sums and products of expectation values for individual spins instead of generating a complicated, shared matrix for the spins which increases exponentially with increasing number of spins. Thus, the presented method is not limited by the system size concerning the number of spins. Instead it is limited by the availability of a quantum chemical method which is able to calculate the electronic ground state for a given configuration $X$ of the nuclei. Thus, the developed method may be applied to molecular systems, which are currently investigated in chemistry, pharmacy, nanoscience or biomedicine. 
\\ \medskip \\
\textbf{Acknowledgements.}
K. T. thanks his wife $\mathfrak{L}_{ t}^\beta$evin, Prof. Jan-Bernd H\"{o}vener and Prof. Klaus Fredenhagen for support and helpful discussions. This work is dedicated to Finn Walter Them.
\\ \medskip \\
\textbf{Author contribution}
K. T. developed most of the present work autonomously before he was employed at UKSH and Kiel University.

\bibliographystyle{apsrev4-1}
\bibliography{DNA-Quantum}

\end{document}